\documentclass[twocolumn, showpacs, preprintnumbers,superscriptaddress, amsmath, amssymb, prb]{revtex4-2}
\usepackage{graphicx}% Include figure files
\usepackage{dcolumn}% Align table columns on decimal point
\usepackage{bm}% bold math
\usepackage{epstopdf}

\usepackage{color}

\begin{document}

\title{
Thermal Broadening of Phonon Spectral Function in Classical Lattice Models: Projective Truncation Approximation }

\author{Hu-Wei Jia}
\affiliation{School of Physics, Renmin University of China, Beijing 100872, China}
\affiliation{Key Laboratory of Quantum State Construction and Manipulation (Ministry of Education), Renmin University of China, Beijing 100872, China}

\author{Wen-Jun Liu}
\affiliation{School of Physics, Renmin University of China, Beijing 100872, China}
\affiliation{Key Laboratory of Quantum State Construction and Manipulation (Ministry of Education), Renmin University of China, Beijing 100872, China}

\author{Yue-Hong Wu}
\affiliation{School of Physics, Renmin University of China, Beijing 100872, China}
\affiliation{Key Laboratory of Quantum State Construction and Manipulation (Ministry of Education), Renmin University of China, Beijing 100872, China}

\author{Kou-Han Ma}
\affiliation{International Center for Quantum Materials and School of Physics, Peking University, Beijing 100871, China}

\author{Lei Wang}
\affiliation{School of Physics, Renmin University of China, Beijing 100872, China}
\affiliation{Key Laboratory of Quantum State Construction and Manipulation (Ministry of Education), Renmin University of China, Beijing 100872, China}

\author{Ning-Hua Tong}
\email{nhtong@ruc.edu.cn}
\affiliation{School of Physics, Renmin University of China, Beijing 100872, China}
\affiliation{Key Laboratory of Quantum State Construction and Manipulation (Ministry of Education), Renmin University of China, Beijing 100872, China}

\date{\today}

\begin{abstract}

Thermal broadening of the quasi-particle peak in the spectral function is an important physical feature in
many statistical systems, but it is difficult to calculate. To tackle this problem, we propose the $H$-expanded basis within the projective truncation approximation (PTA) of the Green's function equation of motion. A zeros-removing technique is introduced to stabilize the iterative solution of the PTA equations. Benchmarking calculations on the classical one-variable anharmonic oscillator model and the one-dimensional $\phi^4$ lattice model show that the thermal broadened quasi-particle peak in the spectral function can be produced on a semi-quantitative level. Using this method, we discuss the low- and high- temperature power-law behaviors of the spectral width $\Gamma_k(T)$ of the one-dimensional $\phi^4$ model, finding it in contradiction with the assumption of effective phonon theory. A short-chain limit of this model is also discovered. Issues of extending the $H$-expanded basis to quantum systems and of the applicability of the Debye formula for thermal conductivity  are discussed. 

\end{abstract}
%\pacs{05.10.Cc, 64.70.Tg, 03.65.Yz, 05.30.Jp}

%\keywords{}

%%%%%%%%%%%%%%%%%%
%%%%%%%%% PACS Possible Selection
%%%%   05.10.Cc Renormalization group methods
%%%%   05.70.Jk Critical point phenomena
%%%%   05.30.Jp Boson systems
%%%%   64.70.Tg Quantum phase transitions
%%%%%%%%%%%%%%%%%%
\maketitle

\begin{section}{Introduction}

The Green's function (GF) is an important tool for studying the physical features of many-body systems in condensed-matter physics.  
The idea of projection  \cite{Mori,Mori2,Zwanzig,Tserkovnikov1,Tserkovnikov2} has been widely used in the calculation of the GF, leading to theories such as the two-pole approximation \cite{Roth1,Roth2}, the composite operator method \cite{Avella1,Haurie1}, the self-consistent projection operator approach \cite{Fulde1}, the operator projection method \cite{Imada1}, continued fraction \cite{Lee1,Lee2}, the irreducible Green's function method \cite{Kuzemsky1}, mode-coupling theory \cite{Lepri1,Castellano1}, and the dynamical projective operatorial approach \cite{Eskandari1}, {\it etc.}. However, in practical applications, complicated analytical derivation and additional uncontrolled approximations are often required in these theories, which hampers systematic improvement of the results. 

Recently, we developed a GF equation of motion (EOM) projective truncation approximation (PTA) method \cite{Fan1,Fan2,Ma1,Fan3}. 
The PTA overcomes the arbitrariness of traditional Tyablicov type truncation \cite{Tyablikov1} by guaranteeing the analytical structure of the GF, diagonal-positivity of the spectral function, and the equilibrium state conservation of energy \cite{Fan1}. Compared to traditional projection-based methods, the PTA avoids the complicated analytical derivations to a large extent, and minimizes the need for additional uncontrolled approximations. 
Within the PTA, the study of a given system is reduced to selecting a basis of operators and solving the generalized eigenvalue problem on that basis. The power of modern computers is thus combined with the idea of projection. By a clever selection of the basis and systematically enlarging its size, exact results can be approached efficiently \cite{Fan2}.

In the present work, we address the issue of describing the damping of quasi particles in the PTA. The damping (decay, relaxation, {\it etc.}) of quasi particles is ubiquitous in the experimental observations of spectra and transport. Modern numerical methods such as quantum Monte Carlo (QMC), density matrix renormalization group (DMRG), and tensor-network renormalization group (TNRG) methods have their own limitations in dealing with the damping. For examples, QMC requires an additional procedure to produce the real frequency spectral function, while DMRG and TNRG are more suitable for low dimensional systems at zero temperature. Theories based on Feynman diagrams usually apply only to weak correlation regime. In the traditional EOM formalism \cite{Tserkovnikov1,Tserkovnikov2} or the generalized Langevin equation approach \cite{Mori,Mori2,Zwanzig}, one needs to evaluate the imaginary part of the self-energy or the memory function to obtain a microscopic description of the damping, which is usually a difficult task. Within the PTA, although the quasi-particle energy can be obtained straightforwardly, the damping is also a challenge. This is because the PTA produces an approximate spectral function composed of $N$ discrete delta peaks, with $N$ being the dimension of the operator basis. Naively increasing $N$ to infinity is shown to be very inefficient for describing the broadening the spectral function and damping of quasi particles. See below.

Here, we propose the PTA on an $H$-expanded basis to solve a part of this problem, {i.e.}, the damping of quasi particles due to the thermal fluctuations at finite temperature. In the spectral function, it corresponds to the thermal broadening of the quasi-particle peak. The same idea can be extended to tackle the quantum damping and this issue will be addressed in the discussion part. In the framework of the PTA, our idea is first to produce the approximate eigen excitation operator $O$, with $[O, H] \approx \epsilon O$, from a low-order PTA. The corresponding spectral function has a delta peak at $\omega = \epsilon$. Then we construct the new basis of operators by expanding $O$ with (exact or approximate) conserved quantities such as the Hamiltonian $H$ of the studied system or the mode Hamiltonian $H_{\nu k}$ of mode $(\nu k)$. That is, we choose the new basis operators in the form $\{ O f_i(H) \}$, with $f_i$ ($i=1, 2, ..., N$) being linearly independent functions. On the level of low-order PTA, these basis operators are degenerate excitation operators corresponding to the excitation energy $\epsilon$. A new PTA calculation on the expanded basis will take into account the thermal fluctuation beyond the low-order PTA and hence split the degeneracy, producing refined spectral functions composed of $N$ closely distributed delta peaks around the low-order quasi-particle peak $\delta(\omega - \epsilon)$. It provides a means to describe the thermal broadening of the quasi particle.

To illustration the implementation and effect of the above idea, we study the phonon damping problem in the classical lattice models which have only thermal damping of quasi particles.
We consider two models. One is the one-variable anharmonic oscillator (AHO) model, whose static averages and spectral function are exactly solvable \cite{Onodera1}. The other is the one-dimensional (1D) $\phi^4$ lattice model which is frequently used in the study of thermal transport in low-dimensional materials. For this model, molecular dynamical (MD) simulation results are available for benchmark.

At finite temperature, the phonon peak in the spectral function of 1D $\phi^4$ lattice will be broadened by thermal fluctuations due to the existence of nonlinear interaction. For general models, various methods have been developed to study the phonon decay and its consequence on the thermal transport, including the effective phonon theory \cite{Li1}, the anharmonic phonons \cite{AP1} and the resonance phonon approaches \cite{Xu1,Xu2} based on MD simulation, non-equilibrium GF \cite{Wang1}, series expansion \cite{Valle1}, mode-coupling theory \cite{Lepri1,Castellano1}, and the Boltzmann equation approach \cite{Nickel1}. But a systematic study of the phonon decay for the 1D classical $\phi^4$ model, in particular, its low- and high-temperature asymptotic behaviors, is still lacking.

Our results show that PTA on the $H$-expanded basis can produce semi-quantitatively accurate $\Gamma_k$, the peak width of the phonon spectral function. It works for all temperatures, including the low temperature limit where MD simulation is too time consuming. Using this method, we find the low- and high-temperature asymptotic power laws in the temperature dependence of $\Gamma_k(T)$ of the 1D $\phi^{4}$ model.

Notably, in contrast to the assumption of the effective phonon theory $\Gamma_k \propto \epsilon \Omega_k$ \cite{Li1}, our results suggest a different relation $\Gamma_k(T) \approx c\left[ \Omega_k(T) - \Omega_k(0) \right]$, with $c$ being weakly dependent on $T$ and momentum $k$. When inserted into the Debye formula \cite{Ziman1,Allen1} for thermal conductivity $\kappa$, it gives different temperature dependence of $\kappa(T)$ from that of MD. This raised question on the applicability of Debye formula for the 1D $\phi^4$ lattice model. We also find a short-chain limit of the $\phi^4$ model with distinct temperature dependence of $\Gamma_k(T)$, which is amenable to experimental test in low dimensional materials. The method could be extended to quantum many-body systems where the damping of quasi particles may occur at zero temperature due to quantum fluctuations. We address this issue in the discussion section.

The rest of this paper is arranged as follows. For completeness, in Sec.II, we first review the formalism of GF EOM and PTA for classical systems. In Sec.III, we develop the formalism of the PTA on an $H$-expanded basis for the AHO model. In Sec.IV, we apply this method to 1D $\phi^4$ lattice model, and we analyze the thermal broadening effect in phonon spectral function. Section V is devoted to a discussion and summary.

\end{section}

\begin{section}{Green's Function EOM and PTA for classical systems}

In this section, we review the formalism of the PTA for classical system, which was developed in Ref.\cite{Ma2}. 
This is to set the stage for developing the PTA on the $H$-expanded basis. 

For two dynamical variables at time $t$ and $t^{\prime}$, $A[q(t), p(t)]$ and $B[q(t^{\prime}), p(t^{\prime})]$ ($q$ and $p$ are abbreviation of canonical coordinates $q_i$'s and momenta $p_i$'s of the studied system), the retarded GF is defined as \cite{Herzel1,Smith1,Cavallo1,Cavallo2}
\begin{equation}       \label{Eq1}
   G^{r}\left[ A(t)|B(t^{\prime}) \right] \equiv \theta(t-t^{\prime}) \langle \{ A(t), B(t^{\prime}) \} \rangle.
\end{equation}
Here $\theta(x)$ is the Heaviside step function. $\langle O \rangle$ is the average of variable $O$ in  the equilibrium state of Hamiltonian $H(q, p)$ at temperature $T$. $\{ \, ,\, \}$ is the Poisson bracket.

The Fourier transformation of the above GF is
\begin{equation}      \label{Eq2}
   G^{r}(A|B)_{\omega} = \int_{-\infty}^{\infty}  G^{r}\left[ A(t)|B(t^{\prime}) \right] e^{i (t-t^{\prime} ) (\omega + i \eta)} d(t-t^{\prime}).
\end{equation}
Here, $\eta$ is an infinitesimal positive number. By making derivative to time $t$ and $t^{\prime}$, we obtain the EOM as \cite{Ma2}
\begin{eqnarray}      \label{Eq3}
   \omega G( A | B )_{\omega} &=& i \langle \{ A, B \} \rangle + i G( \{ A, H \} | B )_{\omega}  \nonumber \\
   &=&i \langle \{ A, B \} \rangle - i G( A| \{ B, H \} )_{\omega}.
\end{eqnarray}
Here, the Zubarev GF $G(A|B)_{\omega}$ without superscript $r$ is related to the retarded GF $G^{r}(A|B)_{\omega}$ by $G^{r}(A|B)_{\omega} = G(A|B)_{\omega + i\eta}$. 

The average value $\langle B A \rangle$ can be obtained from GF through the fluctuation-dissipation theorem \cite{Herzel1,Cavallo2}
\begin{equation}      \label{Eq4}
   \langle B A \rangle = \frac{1}{\beta} \int_{-\infty}^{\infty} \frac{\rho_{A,B}(\omega)}{\omega} d\omega + \langle B_0 A_0 \rangle.
\end{equation}
Here, $A_0$ and $B_0$ are the zero-frequency components of $A(t)$ and $B(t)$, respectively \cite{Ma2}. The spectral function $\rho_{A,B}(\omega)$ in the above equation is defined as
\begin{equation}      \label{Eq5}
   \rho_{A,B}(\omega) = \frac{i}{2\pi} \left[ G( A|B )_{\omega+ i\eta} - G( A|B)_{\omega-i\eta} \right].
\end{equation}

To solve the EOM of GF by the PTA, we first choose a set of linearly-independent basis variables $A_1, A_2, ..., A_N$, which span an $N$-dimensional subspace of variables. Due to the symmetry of the Hamiltonian of the type $H = \sum_i p_i^2/2\mu_i + V(q)$, we can select the basis in one of the subspaces below, 
\begin{eqnarray}    \label{Eq6}
&&  \{O_{e}\} = \{ f(q) \prod_i p_i^{m_i} \,\, \Big{|} \,\, \sum_i m_i=2k, \,\, k \in \mathbb{Z} \}, \nonumber \\
&& \{O_{o}\} = \{ g(q) \prod_i p_i^{n_i} \,\, \Big{|} \,\,\sum_i n_i=2k+1, \,\, k \in \mathbb{Z}\}.\nonumber \\
&&
\end{eqnarray}
Here, $f(q)$ and $g(q)$ are arbitrary functions of $q$. One can prove that the GF matrix $G(\vec{A}|\vec{A})_{\omega}$, defined for the vector of variables $\vec{A} = \left( A_1, A_2, ..., A_N \right)^{T}$, fulfils the EOM
\begin{eqnarray}    \label{Eq7}
&&    \omega^2 G( \vec{A} \big| \vec{A}^{\dag} )_{\omega} =  -\langle \{\{ \vec{A}, H \} , \vec{A}^{\dag}\} \rangle - G( \{ \{\vec{A}, H \}, H \} \big| \vec{A}^{\dag} )_{\omega}. \nonumber \\
&&
\end{eqnarray}

To truncate the EOM in Eq.(\ref{Eq7}), we introduce the inner product of two classical variables $A$ and $B$ as
\begin{equation}    \label{Eq8}
 (A|B) \equiv  \langle \{ A^{\ast}, \{B, H\} \} \rangle.
\end{equation}
We then truncate the EOM as
\begin{equation}    \label{Eq9}
\{ \{ \vec{A}, H\}, H \} \approx - {\bf M}^{T} \left( \vec{A} - \vec{A}_0 \right),
\end{equation}
Projecting this equation to each basis operator $A_k$, one obtains the matrix equation ${\bf M }= {\bf I}^{-1} {\bf L}$. Here ${\bf I}$ is the inner product matrix, with element $I_{ij} \equiv (A_i | A_j)$. ${\bf L}$ is the Liouville matrix, with element $L_{ij} = - (A_i | \{ \{A_j, H\}, H\} )$. Both are Hermitian and positive semi-definite matrices. ${\bf M}$ is thus guaranteed to have real positive eigenvalues. The GF matrix is formally expressed as
\begin{equation}    \label{Eq10}
G( \vec{A}\big| \vec{A^{\dag}} )_{\omega} \approx \left( \omega^2 - {\bf M}^{T} \right)^{-1} {\bf I}^{T}. 
\end{equation}
The spectral function obtained from Eqs.(\ref{Eq5}) and (\ref{Eq10}) reads
\begin{eqnarray}    \label{Eq11}
&& \rho_{A_i, A_j^{\ast}} (\omega) \nonumber \\
&& \approx \sum_{k} \frac{\left( {\bf IU} \right)^{\ast}_{ik} \left( {\bf IU} \right)_{jk}}{2\sqrt{\epsilon_k}} \left[ \delta \left(\omega- \sqrt{\epsilon_k} \right) - \delta \left(\omega + \sqrt{\epsilon_k} \right) \right].   \nonumber \\
&&
\end{eqnarray}
In this equation, ${\bf U}$ and ${\bf \Lambda} = \text{diag}(\epsilon_1, \epsilon_2, ..., \epsilon_N)$ are eigenvector and eigenvalue matrices of ${\bf M}$, respectively. That is, ${\bf U}^{-1}{\bf M}{\bf U} = {\bf \Lambda}$. The eigenvalues $\epsilon_k$'s are real and $\epsilon_k \geqslant 0$ for all $k$. They can be obtained by solving the generalized eigen-value problem,
\begin{equation}    \label{Eq12}
   {\bf LU} = {\bf IU \Lambda}.
\end{equation}
${\bf U}$ can be made to satisfy the generalized unitary condition ${\bf U}^{\dagger} {\bf I} {\bf U} = {\bf 1}$.

From the above expression for the spectral function, Eq.(\ref{Eq4}) gives the averages $C_{ij} \equiv \langle A_i^{\ast}A_j \rangle$ as
\begin{equation}    \label{Eq13}
   {\bf C} \approx  \frac{1}{\beta} {\bf I}{\bf L}^{-1} {\bf I} + {\bf C}_0.
\end{equation}    
Here $(C_0)_{ij} = \langle A_{i0}^{\ast}A_{j0} \rangle$ is the correlation of the static component of variables.
This equation, once being used to calculate the averages appearing in ${\bf I}$ and ${\bf L}$, closes the set of non-linear equations for the averages. The PTA solution to the averages can be obtained by self-consistently solving this set of equations. PTA solution to GF and spectral function are then obtained from Eqs.(\ref{Eq10}) and (\ref{Eq11}).

\end{section}

\begin{section}{One-Variable Anharmonic Oscillator Model}

In this section, the idea of expanding the basis with $H$ to describe the thermal broadening of quasi-particle peak is applied to a simple classical statistical model, the one-variable AHO model.
Its Hamiltonian reads 
\begin{eqnarray}   \label{Vx}
   && H = \frac{1}{2 \mu} p^2 + V(x),   \nonumber \\
   && V(x) = \frac{1}{2} \mu \omega_{0}^2 x^2 + \alpha x^4.
\end{eqnarray}
This model describes the motion of a single oscillator of mass $\mu$ in a quadratic potential with the basic frequency $\omega_0$ and an additional quartic potential of strength $\alpha$. The spectral function $\rho_{x,x}(\omega)$ of variable $x$ in the equilibrium state was solved exactly in the canonical ensemble \cite{Onodera1}.
This is an ideal model for demonstrating the applicability of the PTA on the $H$-expanded basis.
At finite temperature, the spectral function $\rho_{x,x}(\omega)$ has a peak around the effective oscillating frequency $\Omega_p$ with width $\Gamma$. The broadening of the spectral peak can be understood as follows. When the nonlinear potential is present, the oscillating frequency of an oscillator depends on its initial energy $E_0$. In the equilibrium state of a finite temperature $T$, $E_0$ has a thermal fluctuation of order $T$ in the canonical ensemble. This leads to an extended distribution of the oscillating frequencies, and subsequently to the broadening of the spectral peak.

%--------------------------------------------------%
\begin{subsection}{ basis $B_1 \equiv \{x^{2m-1}p^{2n-2}\}$}

Before we present the main subject of this paper, {\it i.e.}, PTA calculation of the spectral function on the $H$-expanded basis, let us first explore the convergence behavior of the spectral function from an ordinary basis, {\it i.e.}, the basis composed of powers of basic variables $x$ and $p$. The results will be contrasted with the PTA results from the $H$-expanded basis.

If we start from the GF $G_{x,x}(\omega)$ and carry out the EOM an even number of times, we will get the higher order GFs $G_{x^{2m-1}p^{2n-2}, x}(\omega)$ ($m,n=1,2,...$). The specific form of the generated variable $x^{2m-1}p^{2n-2}$ is due to the parity symmetry of the potential $V(-x)=V(x)$. We therefore define our basis variable as
\begin{equation}
    A_{mn} =x^{2m-1}p^{2n-2},  \,\,\,\, (m = 1, 2,..., M; \, n = 1,2,..., N)
\end{equation}
They form the basis $B_1 = \{ A_{mn}\}$ ($m = 1, 2,..., M; \, n = 1,2,..., N$), which span a $D = M \times N$ dimensional subspace. Due to the symmetry $V(-x)=V(x)$, the static components of these variables are zero, $\left( A_{mn}\right)_0 = 0$. 

Following the standard formalism of PTA summarized in Sec.II, on this basis we have the inner product matrix
\begin{eqnarray}
 I_{rs, mn} &\equiv & \left(A_{rs} | A_{mn} \right)   \nonumber \\
    & = & c_1 \langle x^{2r+2m-4}\rangle \langle p^{2s+2n-4}\rangle  \nonumber \\
    && + c_2  \langle x^{2r+2m-3} V^{\prime}(x)\rangle \langle p^{2s+2n-6}\rangle   \nonumber \\
    && + c_3 \langle x^{2r+2m-2} V^{\prime\prime}(x) \rangle \langle p^{2s+2n-6}\rangle,
\end{eqnarray}
with $c_1 = [(2m-1)/\mu] \left[ (2r-1)(2n-1) -(2s-2)(2m-2)\right]$, $c_2 = -(2n-2)\left[(2r-1)(2n-3) -(2s-2)(2m-1)\right]$, and $c_3 = (2n-2)(2s-2)$.
The Liouville matrix is obtained as
\begin{eqnarray}
    L_{rs, mn} &=& - \frac{1}{\mu^{2}}(2m-1)(2m-2) \, I_{rs, m-1  \, n+1}  \nonumber \\
               & & + \omega_0^{2} \left[ (2m-1)(4n-3)+2n-2 \right]  \, I_{rs, mn}  \nonumber \\
               && + \frac{4 \alpha}{\mu} \left[ (2m-1)(4n-3)+ 3(2n-2) \right] \, I_{rs, m+1 \,  n} \nonumber\\
               && - \mu^{2} \omega_0^{4} (2n-2)(2n-3) \,  I_{rs, m+1  \, n-1} \nonumber\\
               && - 16 \alpha^{2} (2n-2)(2n-3)  \, I_{rs, m+3 \,  n-1} \nonumber\\               
               && - 8 \alpha \mu \omega_0^{2} (2n-2)(2n-3)  \, I_{rs, m+2  \, n-1}.
\end{eqnarray}
The calculation of matrices ${\bf I}$ and ${\bf L}$ requires evaluating the averages of the type $\langle x^{2k} \rangle$ and $\langle p^{2k} \rangle$. The latter can be obtained exactly from the Gaussian integral, which gives $\langle p^{2k} \rangle = (2k-1)!! (\mu T)^{k}$ ($k= 1, 2, ...$). $\langle x^{2k} \rangle$ needs to be computed self-consistently from GFs. However, since our purpose is to check the convergence behavior of the spectral function on this basis, we use the numerically exact value of $\langle x^{2k} \rangle$. 

From ${\bf I}$ and ${\bf L}$, we can calculate the spectral function $\rho_{x,x}(\omega)$ from the following formalism,
\begin{equation}    \label{Eq18}
 \rho_{x,x}(\omega) = \sum_{k} \frac{ ( \bf{IU})_{1k}^{2} }{ 2 \sqrt{\epsilon_{k} } } \left[ \delta(\omega - \sqrt{\epsilon_{k}})  - \delta(\omega + \sqrt{\epsilon_{k}} ) \right].
\end{equation}
In the above equation, the matrix $\bf{U}$ and the excitation energy $\epsilon_k$'s are determined by the generalized eigen-value problem
\begin{equation}    \label{Eq19}
    \bf{L U} = \bf{I U \Lambda},
\end{equation}
where $\bf{U}$ is the generalized eigen vector matrix, and ${\bf \Lambda} = \text{diag}(\epsilon_1, \epsilon_2, ..., \epsilon_D)$ is the generalized eigen value matrix.

\end{subsection}
%---------------------------------------------------%

%---------------------------------------------------%
\begin{subsection}{ $H$-expanded basis $B_2 \equiv \{xe^{\lambda_i H} \}$ and $B_3 \equiv \{x, x^3 \} \otimes \{e^{\lambda_i H} \} $}
% 
% --- Formula: <<Z8>> Eq.(62)-(73). ---
% 
In this subsection, we implement PTA on two successively larger $H$-expanded bases, $B_2 \equiv \{xe^{\lambda_i H} \}$ and $B_3 \equiv \{x, x^3 \} \otimes \{e^{\lambda_i H} \} $ ($i=1, 2, ..., N$). Here $\{ \lambda_i \}$ are $N$ different real numbers. We choose the exponential function $e^{\lambda_i H}$ to expand the low order basis variables $\{ x\}$ and $\{x, x^3\}$, because with this functional form, the calculation of ${\bf I}$ and ${\bf L}$ can be simplified significantly. For details, see Appendix A. The PTA results from bases $B_2$ and $B_3$ will be compared with the exact ones.

\begin{subsubsection}{ basis $B_2 \equiv \{xe^{\lambda_i H} \}$ }

We first derive PTA equations based on the $H$-expanded basis $B_2$ composed of variables
\begin{equation}
    A_{i} \equiv x f_i, \,\,\,\,\, (i=1, 2, ..., N).
\end{equation}
Here, $f_i = e^{\lambda_i H}$. To put the physically interesting variable $x$ into the basis, we assign $\lambda_{1} \equiv 0$, meaning $A_1 = x$. $\{ A_i \}$ form a $N$-dimensional 
basis, with zero static components $A_{i0} = 0$. The parametrization of $\{ \lambda_i \}$ will be specified in Sec.III.D.

The inner product matrix $\bf{I}$ and Liouville matrix $\bf{L}$ are derived as
\begin{eqnarray}   \label{Eq21}
 && I_{ij} =  \frac{\beta}{\mu (\beta - \lambda_i - \lambda_j)} \langle f_i f_j \rangle,  \nonumber \\
 && L_{ij} = \frac{\beta}{\mu^2 (\beta - \lambda_i - \lambda_j) } \left[ \mu \omega_0^{2} \langle f_if_j \rangle + 12 \alpha \langle x^2 f_i f_j \rangle  \right].   \nonumber \\
 &&
\end{eqnarray}
Both ${\bf I}$ and ${\bf L}$ contain the correlation among the expanding factors $\{ e^{\lambda_i H} \}$ that takes into account the energy fluctuation effect. Details of the derivation are given in Appendix A.

To evaluate the averages appearing in $I_{ij}$ and $L_{ij}$, we first note that $\langle x^2 f_i f_j \rangle $ in $L_{ij}$ can be expressed in terms of $C_{ij} \equiv \langle A_i A_j \rangle$ and thus can be calculated self-consistently from the fluctuation-dissipation theorem Eq.(\ref{Eq13}) as
\begin{eqnarray}   \label{Eq22}
   C_{ij} &=& \langle x^2 f_i f_j \rangle  \nonumber \\
   &=& \frac{1}{\beta} \left( {\bf I} {\bf L}^{-1} {\bf I} \right)_{ij}.
\end{eqnarray}
Here, ${\bf C}_0 = 0$ has been used. In particular, we have $\langle x^2 \rangle = C_{11}$.
The other unknown average $\langle f_i f_j \rangle$ cannot be written in the form $\langle A_i A_j \rangle$. For this specific model, it can be evaluated by numerical integration, or by a quadratic variational approximation \cite{Dauxois1}. We have compared the two methods and find little difference in the final results. Here we use the latter method,
\begin{eqnarray}   \label{Eq23}
  \langle f_i f_j \rangle &\approx & \langle f_i f_j \rangle_{vari}  \nonumber \\
 & =& \frac{\int\int_{-\infty}^{\infty} dx dp \, e^{- (\beta-\lambda_i - \lambda_j) H_{vari}} }{ \int\int_{-\infty}^{\infty} dx dp \, e^{-\beta H_{vari}} }.
\end{eqnarray}
The variational Hamiltonian $H_{vari}$ is obtained from a quadratic variational approximation as
\begin{equation}   \label{Eq24}
 H_{vari} = \frac{1}{2 \mu} p^2 + \theta x^2,
\end{equation}
with $\theta = \frac{1}{4} \mu \omega_{0}^{2} + \sqrt{ \left(\frac{1}{4} \mu \omega_{0}^{2} \right)^2 + 3 \alpha T}$.
Using this variational Hamiltonian, it is easy to obtain the expression for $\langle f_i f_j \rangle$ as
\begin{equation}   \label{Eq25}
 \langle f_i f_j \rangle \approx \frac{\beta}{\beta - \lambda_i - \lambda_j},
\end{equation}
being independent of the parameters in $H$.

The spectral function $\rho_{x,x}(\omega)$ is then calculated from Eqs.(\ref{Eq18}) and (\ref{Eq19}). 

\end{subsubsection}

\begin{subsubsection}{ basis $B_3 \equiv \{x, x^3 \} \otimes \{e^{\lambda_i H} \} $}

Here, we derive the ${\bf I}$ and ${\bf L}$ matrices on basis $B_3$. We denote the basis operator as
\begin{equation}   \label{Eq26}
   A_{\mu i} = a_{\mu} f_i, \,\,\,\, (\mu = 1,2; \,\, i = 1,2,...,N).
\end{equation}
where $a_1 = x$ and $a_2 = x^3$, and $f_{i} = e^{\lambda_{i} H}$ ($i=1, 2, ..., N$).
These variables also have zero static components $\left( A_{\mu i} \right)_0 = 0$. 
The basis $B_3 \equiv \{a_{\mu}f_i \}$ ($\mu = 1,2; \,\, i = 1,2,...,N$) is a $2N$ dimensional subspace, being larger than $B_2$. 

We denote the matrix elements of ${\bf I}$ and ${\bf L}$ as $I_{\mu \nu}^{ij} = \left(A_{\mu i}| A_{\nu j} \right)$, and $L_{\mu \nu}^{ij} = -\left( A_{\mu i}| \{ \{ A_{\nu j}, H\}, H\} \right)$, respectively.
Using the same method for deriving ${\bf I}$ and ${\bf L}$ on the $B_2$ basis and some additional simplifications, we obtain
\begin{eqnarray}   \label{Eq27}
&&{\bf I}^{ij} = \frac{\beta}{\mu\left(\beta - \lambda_i - \lambda_j \right)}\left( 
\begin{array}{cc} 
\langle f_i f_j \rangle & 3 \langle x^2 f_i f_j \rangle \\ 
3\langle x^2 f_i f_j \rangle & 9 \langle x^4 f_i f_j \rangle \\
\end{array}
\right).    \nonumber \\
&&
\end{eqnarray}
The matrix elements of ${\bf L }$ read
\begin{eqnarray}   \label{Eq28}
L_{11}^{ij} &=& \frac{\beta}{\mu^{2} \left( \beta - \lambda_i - \lambda_j\right)} \left[ \mu \omega_0^2 \langle f_i f_j \rangle + 12 \alpha \langle x^2 f_i f_j \rangle \right],  \nonumber \\
L_{12}^{ij} &=& L_{21}^{ij}    \nonumber \\
&=& \frac{3 \beta}{\mu^{2} \left( \beta - \lambda_i - \lambda_j \right)} \left[ \mu \omega_0^2 \langle x^2 f_i f_j \rangle + 12 \alpha \langle x^4 f_i f_j \rangle \right],  \nonumber \\
 L_{22}^{ij} &=& -\frac{54 \beta}{ \mu^{2} \left(\beta - \lambda_i - \lambda_j \right)^2 } \langle x^2 f_i f_j \rangle   \nonumber \\
 && + \frac{\beta }{\beta - \lambda_i - \lambda_j } \left[ \frac{63 \omega_0^2}{\mu}\langle x^4 f_i f_j \rangle + \frac{324 \alpha}{\mu^2} \langle x^6 f_i f_j \rangle\right].   \nonumber \\
 &&
\end{eqnarray}
Details of the derivation are given in Appendix A.

$C_{\mu \nu}^{ij} = \langle a_{\mu} a_{\nu}f_i f_j \rangle$ is still given by Eq.(\ref{Eq13}), with ${\bf C}_0 = 0$. Some unknown averages in ${\bf I}$ and ${\bf L}$ can be calculated self-consistently through
\begin{eqnarray}
  && \langle x^2 f_i f_j \rangle = C_{11}^{ij}, \nonumber \\
  && \langle x^4 f_i f_j \rangle = C_{12}^{ij}, \nonumber \\
  && \langle x^6 f_i f_j \rangle = C_{22}^{ij}.    
\end{eqnarray}
The other quantities, including $\langle f_i f_j \rangle$, GF, and the spectral function $\rho_{x,x}(\omega)$, can be calculated similarly as for basis $B_2$.

\end{subsubsection}

\end{subsection}
%--------------------------------------------------%

%--------------------------------------------------%
\begin{subsection}{ removing the redundant basis variables }

In our numerical solution of the above PTA equations, we supplemented a technique that we called the zeros-removing method to stabilize the iteration process. It will be useful in general PTA calculation. Therefore, in this subsection, we introduce this method in the general framework of PTA.

In principle, the general PTA formalism works for arbitrary basis dimension $N$. In practice, however, if the PTA calculation is implemented naively, it is limited to some dimension $N$ by numerical instability. This is because the orthogonality of the basis operators are difficult to control and they easily become linearly dependent when $N$ is large. As a result, the inner product matrix $\bf{I}$, which should be positive-definite in theory, often becomes nearly singular for large $N$. This makes the iterative process unstable. This problem, arising from the non-orthogonality of the basis variables, is quite common in PTA calculations. For example, in our study of Anderson impurity model with the PTA \cite{Fan1}, the singularity of $\bf{I}$ also appears.

To partly overcome this difficulty, here we propose a method commonly used in the ab initio calculation with unorthogonal basis \cite{Ren1}. The idea of the method is to first figure out the redundant basis variables by diagonalizing ${\bf I}$ and find certain smallest eigen values and the associated eigen vectors. Then one removes these vectors from the basis and redoes the PTA calculation on the remaining basis. Our calculation shows that this method can significantly stabilize the iteration and makes it possible to do PTA calculation on a much larger basis. Below, we present the formalism of this method.

Suppose we have a set of basis variables $\{ A_i \}$ ($i =1, 2, ..., N$).
When $\bf{I}$ and $\bf{L}$ matrices are obtained, we first diagonalize $\bf{I}$ using the unitary transformation,
\begin{equation}
  ({\bf U}_{I})^{-1} {\bf I} {\bf U}_{I} = {\bf \Lambda}_{I}.
\end{equation}
Here ${\bf U}_{I}$ is an unitary matrix and ${\bf \Lambda}_{I}$ is diagonal.
Among the $N$ eigen values of ${\bf I}$, suppose the smallest $N_1$ of them are close to zero, with their absolute value below a criterion, say, $10^{-12}$. The other $N_2 = N - N_1$ eigenvalues $\{ i_1, i_2, ..., i_{N_2} \}$ are larger than this criterion. Then we can approximately write the matrix ${\bf \Lambda}_{I}$  in the following block form,
\begin{equation}
  {\bf \Lambda}_{I} \approx \left(
\begin{array}{cc}
{\bf 0} &  {\bf 0}  \\
{\bf 0} &  {\bf \Lambda}_{I2}  \\
\end{array}
\right),
\end{equation}
with ${\bf \Lambda}_{I2} = {\text{diag}}(i_1, i_2, ..., i_{N_2})$ and $i_k >0$ ($k=1,2,..., N_2$).
We can also write ${\bf U}_{I}$ as
\begin{equation}
{\bf U}_{I} = \left(
\begin{array}{cc}
{\bf U}_{I1} &  {\bf U}_{I2} 
\end{array}
\right),
\end{equation}
where $\left( {\bf U}_{I1} \right)_{N \times N_1}$ contains the $N_1$ column vectors of the zero-length variables, while $\left( {\bf U}_{I2} \right)_{N \times N_2}$ contains the $N_2$ column vectors of finite-length variables. 
From the orthonormal and complete conditions ${\bf U}_{I}^{\dagger} {\bf U}_{I} = {\bf U}_{I} {\bf U}_{I}^{\dagger} = {\bf 1}$, we have the following properties for ${\bf U}_{I1}$ and ${\bf U}_{I2}$,
\begin{eqnarray}
 && {\bf U}_{I1} {\bf U}_{I1}^{\dagger} + {\bf U}_{I2} {\bf U}_{I2}^{\dagger} = {\bf 1}_{N \times N}, \nonumber \\
 && {\bf U}_{I1}^{\dagger} {\bf U}_{I1} = {\bf 1}_{N_1 \times N_1}, \,\,\,\,\,\,\,\,  {\bf U}_{I2}^{\dagger} {\bf U}_{I2} = {\bf 1}_{N_2 \times N_2},  \nonumber \\
 && {\bf U}_{I1}^{\dagger} {\bf U}_{I2} = {\bf 0}_{N_1 \times N_2}, \,\,\,\,\,\,\,\,  {\bf U}_{I2}^{\dagger} {\bf U}_{I1} = {\bf 0}_{N_2 \times N_1}.
\end{eqnarray}

We normalize the finite-length variables by defining matrix $\tilde{\bf U}_{2} = {\bf U}_{I2} \frac{1}{\sqrt{{\bf \Lambda}_{I2}}}$. 
Now we can write down the orthogonalized variables. The zero-length variables are denoted as $O_{1k}$ and the normalized finite-length variables as $O_{2k}$. They read
\begin{eqnarray}
&&   O_{1k} = \sum_{i=1}^{N} A_{i} \left( U_{I1} \right)_{ik},  \nonumber \\
&&   O_{2k} = \sum_{i=1}^{N} A_{i} \left( \tilde{U}_{2} \right)_{ik}.
\end{eqnarray}
Since $ O_{1k}$ has zero length, one can prove that the GF involving $O_{1k}$ are zero, {\it i.e.}, $ G(O_{1k}|O_{1p})_{\omega} = G(O_{1k}|O_{2p})_{\omega} = 0$. Its static correlation function is also zero, {\it i.e.}, $\langle O_{1k} X \rangle =0$ for arbitrary variable $X$. We can therefore focus only on the remaining $N_2$-dimensional subspace.
The original basis variable $A_i$ is now expressed in terms of $O_{1k}$ and $O_{2k}$ as
\begin{equation}    \label{Eq37}
   A_i = \sum_{k=1}^{N_1} O_{1k} \left( {\bf U}_{I1}^{\dagger} \right)_{ki} + \sum_{k=1}^{N_2} O_{2k} \left( {\bf \Lambda}_{I2} \tilde{\bf U}_{2}^{\dagger} \right)_{ki}.
\end{equation}

The Liouville matrix on the normalized basis $O_{2k}$ reads
\begin{equation}
   \tilde{\bf L}_{22} = \tilde{\bf U}_{2}^{\dagger}  {\bf L} \tilde{\bf U}_{2}.
\end{equation}
Here $\tilde{\bf L}_{22}$ is a $N_2 \times N_2$ Hermitian matrix. We diagonalize it by a unitary transformation
\begin{equation}
    [{\bf V}_{22} ]^{-1} \tilde{\bf L}_{22} {\bf V}_{22} = {\bf \Lambda}_{2},
\end{equation}
with ${\bf \Lambda}_{2} = \text{diag}(\epsilon_1, \epsilon_2, ..., \epsilon_{N_2})$ being the square of the $N_2$ eigen excitation energies.

In the $N_2$-dimensional space of the normalized basis variables, we employ the standard PTA to write down the GF matrix as
\begin{eqnarray}
   \tilde{\bf G}_{22}(\omega) &=& \left( \omega^2 {\bf 1} - \tilde{\bf L}_{22}^{T} \right)^{-1} \nonumber \\
     &=& \left({\bf V}_{22}^{-1} \right)^{T} \left(\omega^2 {\bf 1} - {\bf \Lambda}_2 \right)^{-1} {\bf V}_{22}^{T}. 
\end{eqnarray}
In the above equation, $(\tilde{G}_{22} )_{kp}(\omega) = G(O_{2k} | O_{2p}^{\ast})_{\omega}$.
Using Eq.(\ref{Eq37}), we obtain the GF of original basis variables $A_i$ as
\begin{equation}    \label{Eq41}
G(A_i|A_j^{\ast})_{\omega} = \sum_{k,p=1}^{N_2} \sqrt{i_k i_p} \left( U_{I2}\right)^{\ast}_{ik}\left( U_{I2} \right)_{jp}
( \tilde{G}_{22} )_{kp}(\omega).
\end{equation}
The spectral function is obtained similarly as
\begin{eqnarray}     \label{Eq42}
&&  \tilde{\rho}_{O_{2k}, O_{2p}^{\ast}}(\omega) \nonumber \\
&& = \sum_{m=1}^{N_2} \frac{ \left(V_{22}\right)^{\ast}_{km}\left(V_{22}\right)_{pm}}{2 \sqrt{\epsilon_m}} \left[ \delta(\omega - \sqrt{\epsilon_m}) - \delta(\omega + \sqrt{\epsilon_m})  \right],   \nonumber \\
&& 
\end{eqnarray}
and
\begin{equation}  \label{Eq43}
\rho_{A_i, A_j^{\ast}}(\omega) = \sum_{k,p=1}^{N_2} \sqrt{i_k i_p} \left( U_{I2}\right)^{\ast}_{ik}\left( U_{I2} \right)_{jp}  \tilde{\rho}_{O_{2k}, O_{2p}^{\ast}}(\omega).
\end{equation}
From the spectral theorem, the averages $( \tilde{C}_{22} )_{kp} = \langle O_{2k}^{\ast} O_{2p}\rangle$ are calculated as
\begin{equation}
  \tilde{\bf C}_{22} = \frac{1}{\beta} \left( \tilde{\bf L}_{22} \right)^{-1}.
\end{equation}
The correlation function between the original basis variables reads
\begin{equation}
   \langle A_{j}^{\ast} A_i \rangle = \sum_{k,p=1}^{N_2} \sqrt{i_k i_p} \left( U_{I2}\right)^{\ast}_{ik}\left( U_{I2} \right)_{jp}  \left[ \tilde{C}_{22} \right]_{pk}.
\end{equation}

The closed equations for $\langle A_j^{\ast} A_i \rangle$ are obtained by expressing ${\bf I}$ and ${\bf L}$ in terms of$\langle A_j^{\ast} A_i \rangle$. They can be solved iteratively. Once $\langle A_j^{\ast} A_i \rangle$ are obtained, the GF and spectral functions can be calculated from Eqs.(\ref{Eq41}) and (\ref{Eq43}). 

\end{subsection}
%---------------------------------------------------%

%---------------------------------------------------%
\begin{subsection}{results for one-variable AHO model}

In this subsection, we present the results for the one-variable AHO model, obtained from the basis $B_{1}\equiv \{ x^{2m-1}p^{2n-2} \}$ ($ 1 \leq m \leq M$, $ 1 \leq n \leq N$), $B_{2} \equiv \{ x e^{\lambda_i H} \}$ ($1 \leq i \leq N$), and $B_{3} \equiv \{ x, x^{3} \} \otimes \{ e^{\lambda_i H} \}$ ($1 \leq i \leq N$). We compare these results with those from exact numerical integration (for averages) and exact formula \cite{Onodera1} (for the spectral function). We want to show that, although the $B_1$ basis becomes complete in the limit $M=N = \infty$, the shape of the primary peak in the spectral function converges very slowly. The $B_2$ basis quite faithfully produces the shape of the main peak with moderate dimension $N$. The $B_3$ basis produces both the main peak and the secondary peak. These results demonstrate that the $H$-expanded basis can efficiently describe the thermal broadening effect in the spectral function. For the data shown below, we typically use the smallness criterion $10^{-14} \sim 10^{-9}$ for truncating the spectrum of ${\bf I}$ and the convergence criterion $10^{-9} \sim 10^{-7}$ for iterative solution of $\langle A_i^{\ast} A_j \rangle$. All our results in this paper are in the unit $k_B=1$.

The $H$-expanded bases $B_2$ and $B_3$ are characterized by $N$ real parameters $\{ \lambda_i \}$ ($i=1,2,...,N$). We find that the value of these parameters will influence the convergence speed of PTA calculation and the quality of the spectral function. Since in $\langle f_if_j \rangle \sim \text{Tr}[ e^{-(\beta - \lambda_i -\lambda_j)}]$, $\lambda_i + \lambda_j  < \beta$ is required, we let $\lambda_i \in (-\infty, \beta/2)$. The lower and higher $\lambda_i$'s are responsible for describing low and high energy thermal fluctuations, respectively. We use the following scheme for assigning them,
\begin{equation}     \label{Eq46}
   \lambda_i = 
 \left\{\begin{array}{lll}   
   \frac{1}{2} \left(\frac{1}{T} - \frac{1}{T_i} \right) & \,\,\, ( 2 \leq i \leq N), \\
        \\ 
0 & \,\,\, (i =1). 
     \end{array} \right. 
\end{equation}
Here, we set $\lambda_1 = 0$ to include $x$ in the basis. We let $\ln{T_i}$ distribute equal-distantly in the interval $[\ln{T} - \Delta, \ln{T}+ \Delta ]$.  That is, for $N > 1$, we use
\begin{equation}    \label{Eq47}
   \ln{T_i} = \ln{T} - \Delta + (i-1) \frac{2 \Delta}{N-1}, \,\, \,\,\,\,\,\,( 2 \leq i \leq N). 
\end{equation}
Typically $\Delta = 4.0 \sim 9.0$ is used in our calculation. 

Our test shows that the above scheme works quite well for the AHO model. The peaks in the spectral function $\rho_{x,x}(\omega)$ shift slightly with $\Delta$ in the range $\Delta = 1.0 \sim 10.0$. The key features extracted from $\rho_{x,x}(\omega)$, such as the peak position $\Omega_p$, peak width $\Gamma$, and the moments of the spectral function, change very little. As $\Delta$ decreases, $\Gamma$ stays qualitatively the same until $\Delta \sim 10^{-6}$, at which point $\Gamma$ jumps to zero, recovering the expected result for $\Delta =0$. For tiny but finite $\Delta$, $\rho_{x,x}(\omega)$ has two delta peaks, since the basis is effectively reduced to a two-dimensional basis $\{x, x H \}$ in the small $\Delta$ limit. The same holds for the 1D $\phi^{4}$ lattice model. When we need to average the spectral function $\rho_{x,x}(\omega)$ over $N_r$ random assignments of $\{ \lambda_i\}$,
we let $\ln{T_i}$ take random numbers from the uniform distribution within $[\ln{T} - \Delta, \ln{T}+ \Delta]$. 

In this work, when presenting the spectral function, we always broaden the delta peaks in the spectral function by a Gaussian function with standard variance $\sigma$. Depending on the occasion, the value of $\sigma$ is either fixed to be $0.2 \sim 0.05$, or it is taken to be proportional to the width of the main peak $\Gamma$, $\sigma = r \Gamma$ with $r = 0.2 \sim 0.3$. 

Considering that the spectral function in the present study always has a single primary peak in the $\omega > 0$ regime, and it is almost symmetric in shape, we extract the peak position $\Omega_p$ of the primary peak through
\begin{equation}    \label{Eq48}
   \Omega_p = m_1,
\end{equation}
and the width $\Gamma$ of the primary peak through
\begin{equation}     \label{Eq49}
\Gamma = \sqrt{ m_2 - m_1^{2}}.
\end{equation}
Here, $m_k$ is the $k$-th order moment of the spectral function on the positive frequency axis. This method was used in, {\it e.g.}, defining the spin wave line width \cite{Winterfeldt1}. $m_k$ is evaluated from
the weight $w_i$ and pole $p_i$ of the raw data of spectral function $\rho_{x,x}(\omega) = \sum_{i} w_i \delta(\omega - p_i)$ as $m_k = (\sum_{i, p_i > 0} w_i p_i^{k} ) / ( \sum_{i, p_i > 0 } w_i )$.  
In this way, both $\Omega_p$ and $\Gamma$ are independent of the broadening procedure used for plotting the curve of $\rho_{x,x}(\omega)$. Note that in the literature, there are various ways for extracting the width of phonon peak from the MD simulation data \cite{AP1,Xu1,Guo1,YLiu1}.

\begin{figure}
 \vspace*{-4.5cm}
\begin{center}
  \includegraphics[width=430pt, height=320pt, angle=0]{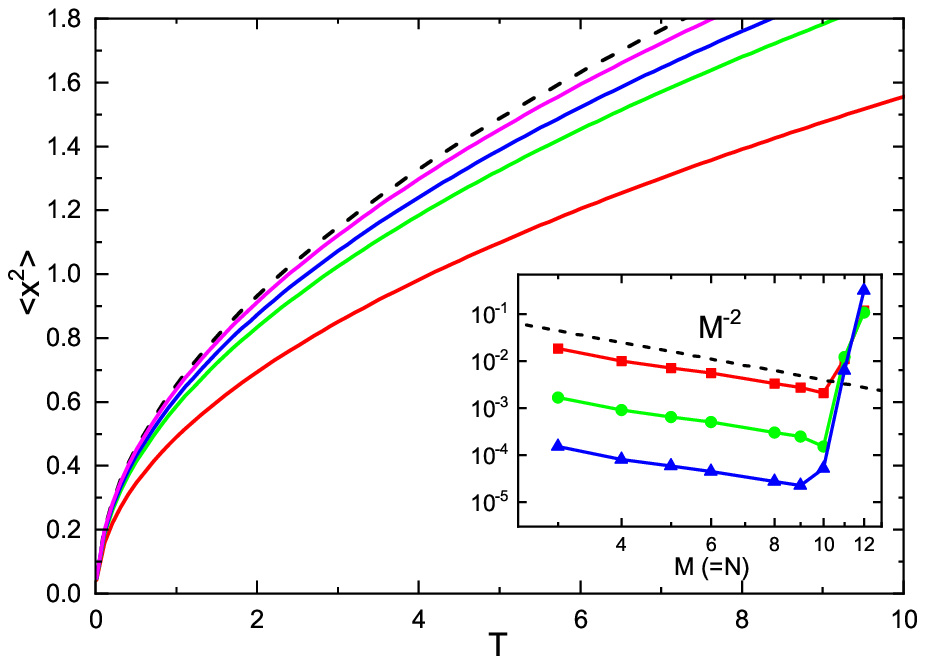}
  \vspace*{-1.0cm}
\end{center}
  \caption{(color online) $\langle x^{2}\rangle$ as functions of $T$ for AHO model. Solid lines are obtained from $B_1$-PTA with $M=1$, $2$, $3$, and $6$ from bottom to top ($N=M$ is always used). The dashed line is for the exact solution. Inset: The errors with respect to the exact results as functions of $M$ ($N=M$) in $\langle x^{2} \rangle$ (red squares), $\langle x^{4} \rangle$ (green dots), and $\langle x^{6} \rangle$ (blue triangles) at $T=0.3$ . The dashed line is $M^{-2}$ for guiding eyes. Model parameters are $\mu =1.0$, $\omega_0 = 0.3$, and $\alpha=0.25$.
}   \label{Fig1}
\end{figure}
\begin{figure}
 \vspace*{-2.5cm}
\begin{center}
  \includegraphics[width=350pt, height=290pt, angle=0]{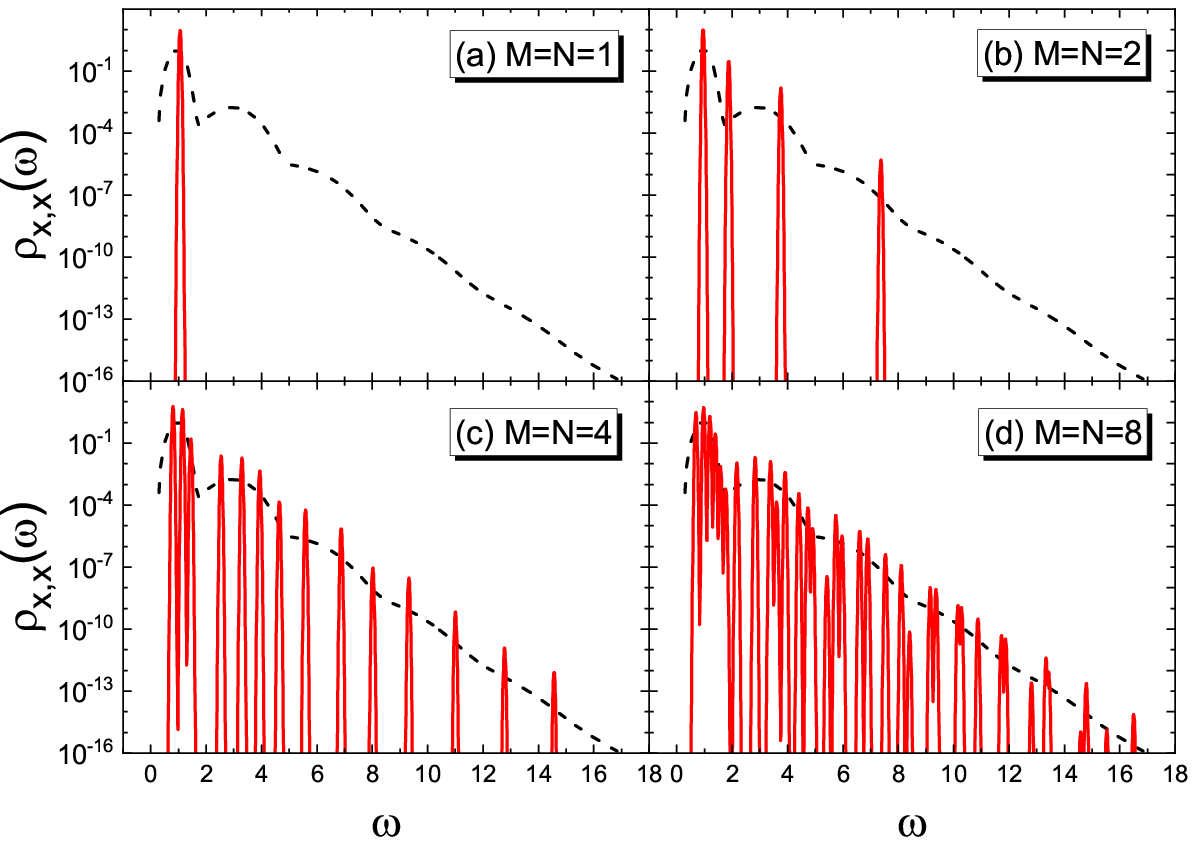}
  \vspace*{-1.0cm}
\end{center}
  \caption{(color online) Comparison of spectral function $\rho_{x,x}(\omega)$ obtained from $B_1$-PTA (red solid line) with the exact curve (black dashed line). Model parameters are $\mu =1.0$, $\omega_0 = 0.3$, $\alpha=0.25$, and $T=0.3$. The delta peaks in the PTA curves are broadened using Gaussian function with standard variance $\sigma=0.02$.
}   \label{Fig2}
\end{figure}

Below, we first examine the convergence behavior of the basis $B_1$. In the limit $M=N=\infty$, $B_1$ becomes complete and the
results from PTA are expected to become exact. In Fig.1, we show the average value $\langle x^2 \rangle$ as a functions of $T$ for various $M$ (fixing $N=M$). It clearly shows that as $M=N$ increases, the curve uniformly approaches the exact dashed line. The inset of Fig.1 shows that the errors of $\langle x^2 \rangle$, $\langle x^4 \rangle$, and $\langle x^6 \rangle$ at $T=0.3$ decrease with increasing $M$ as $M^{-2}$ (fixing $N=M$). It is also observed that at about $M=N=10$, {\it i.e.}, at the dimension of basis $D=M \times N =100$, the error from PTA abruptly increases. We have observed that the largest (smallest) eigenvalue of ${\bf I}$ increases (decreases) exponentially with increasing $M=N$. At about $M=N=10$, the condition number of ${\bf I}$ deteriorates to such an extent that the iterative solution of PTA equations can no longer be carried out accurately. 
In our previous application of PTA \cite{Fan1}, we also observed similar phenomena, hinting that this problem may be universal for PTA. Note that this problem can not be solved by the use of zeros-removing method discussed in Sec.III.C.

In Fig.2, the spectral function $\rho_{x,x}(\omega)$ of AHO model at $T=0.3$ is shown. In each panel, the red curve (from PTA on basis $B_1$) is compared to the black dashed line (from exact solution \cite{Onodera1}). The exact $\rho_{x,x}(\omega)$ has a series of broad peaks at successively higher frequencies, whose weights decay exponentially with frequency. These multiple peaks are the overtone phenomenon due to the non-linear confinement, broadened by thermal fluctuations. At very low temperature, the peak positions tend to frequencies $\omega_0$, $3\omega_0$, $5 \omega_0$, {\it etc.}.
The exact spectral function vanishes abruptly for $\omega < \omega_0$, in agreement with the fact that $\omega_0$ is the lowest oscillation frequency for all initial conditions. 

It is seen that as $M=N$ increases up to $M=N=8$, the PTA-produced spectral function contains more and more peaks (we broaden the delta peaks a little for presentation purpose). The peak positions and weights are distributed in such a way that the envelope curve uniformly approaches the exact one, including the primary peak and the higher order ones. Fig.1 and Fig.2 show that both the thermodynamical and the dynamical quantities indeed converge towards the exact value as the dimension of basis increases.

\begin{figure}
 \vspace*{-3.50cm}
\begin{center}
  \includegraphics[width=420pt, height=300pt, angle=0]{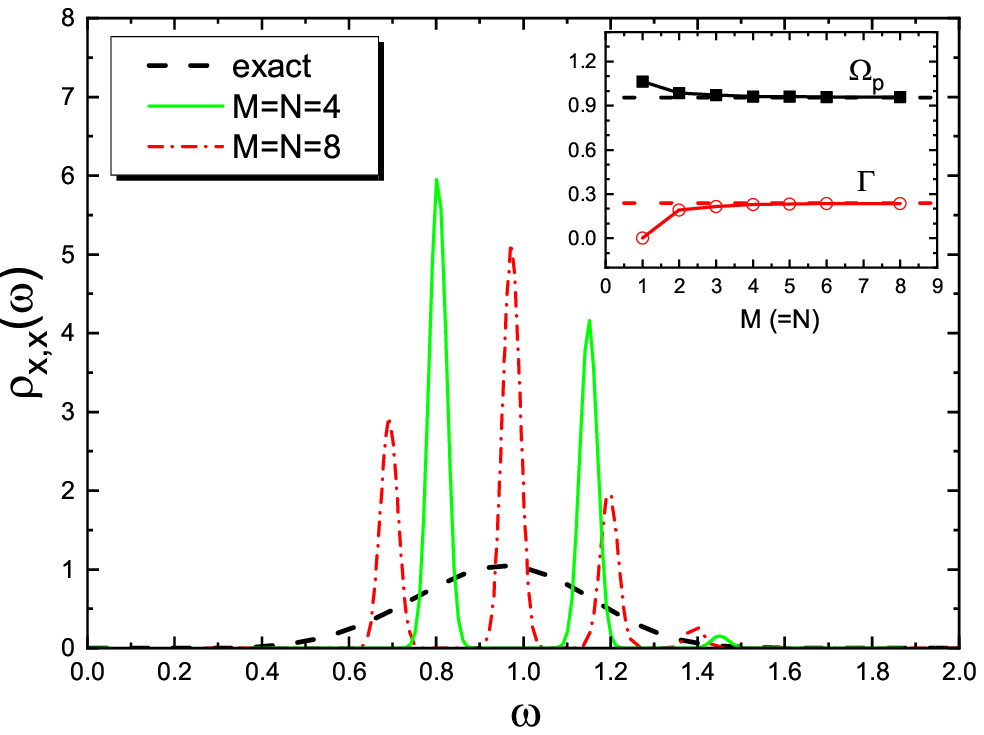}
  \vspace*{-1.0cm}
\end{center}
  \caption{(color online) Primary peak in the spectral function of AHO model, obtained from basis $B_1 = \{x^{2m-1}p^{2n-2} \}$ ($m=1,2,...,M, \,\, n=1, 2, ..., N$) with $M=N=4$ (green solid line) and $M=N=8$ (red dash-dotted line). The black dashed line is the exact curve. Inset: the peak position (black squares) and peak width (read circles) obtained using different $M$ ($N=M$). The dashed lines are exact value. Model parameters are $\mu=1.0$, $\omega_0 = 0.3$, $\alpha=0.25$, and $T=0.3$. For the main figure, we used a constant broadening parameter $\sigma=0.02$.
}   \label{Fig3}
\end{figure}

However, if we look at the primary peak alone in Fig.3 on the linear scale, we will find that with increasing $M=N$, the primary peak of $\rho_{x,x}(\omega)$ from PTA receives little improvement. Up to $M=N=8$, the primary peak has got only $4$ discernible peaks on the linear scale. Clearly, on the naive basis $B_1$, the poles of GF distribute uniformly in the whole frequency axis. Among the $M\times N$ delta peaks, most of them have high frequencies and tiny weights, such that they contribute little to the broad primary peak. This leads to a very slow convergence in the shape of the primary peak with increasing basis dimension, although the extracted peak position $\Omega_p$ and width $\Gamma$ converge fast with increasing $M=N$, as shown in the inset of Fig.3.

\begin{figure}
 \vspace*{-3.5cm}
\begin{center}
  \includegraphics[width=390pt, height=290pt, angle=0]{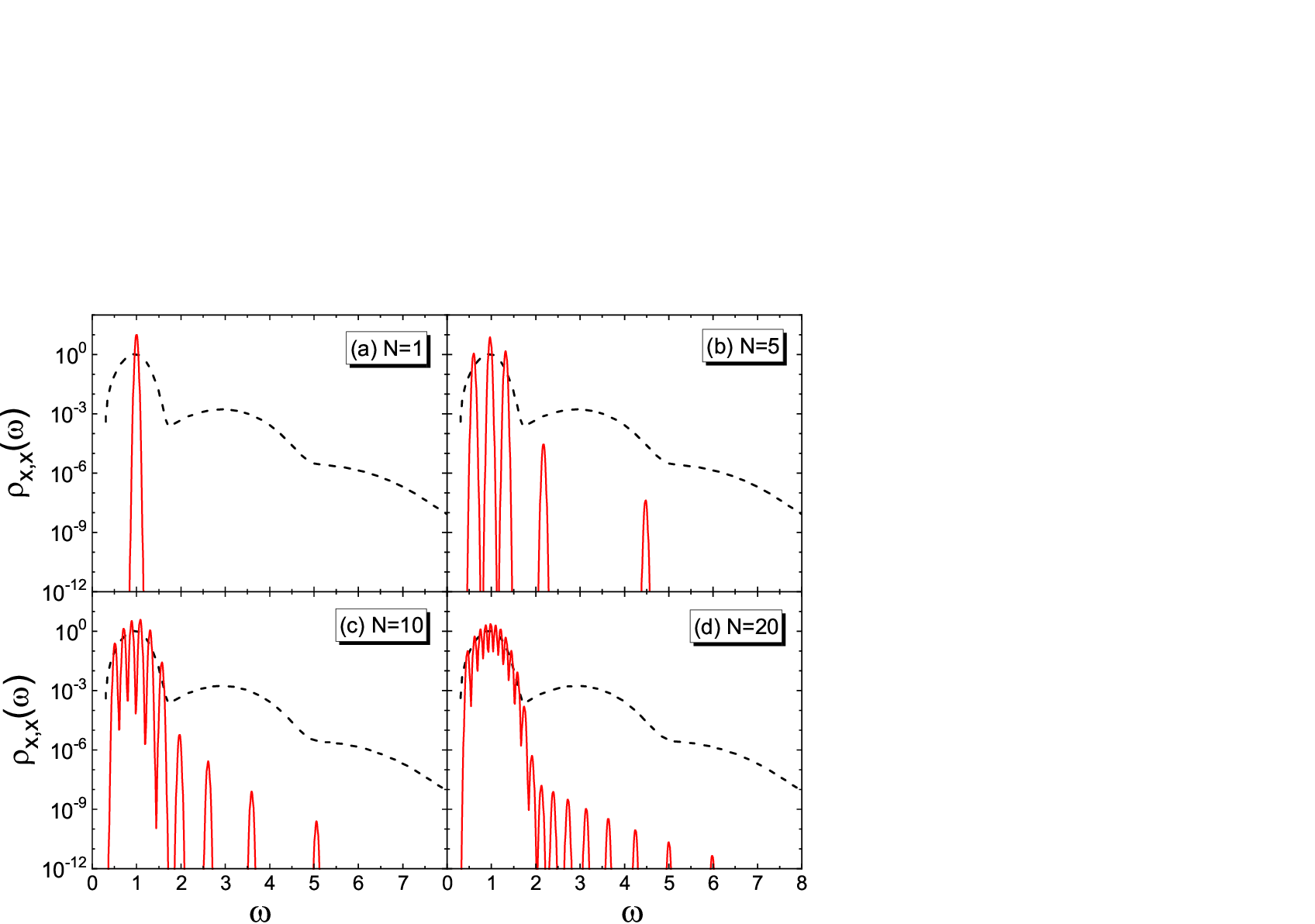}
  \vspace*{-1.0cm}
\end{center}
  \caption{(color online) Comparison of spectral function $\rho_{x,x}(\omega)$ of AHO model, obtained from the $H$-expanded basis $B_2 =\{x e^{\lambda_i H} \}$ ($i=1,2,...,N$) (red solid line) and from the exact solution (black dashed line). 
Model parameters are $T=0.3$, $\mu=1.0$, $\omega_0 = 0.3$, and $\alpha=0.25$. We use $\Delta = 6.0$ and broaden parameter $\sigma = 0.02$.
}   \label{Fig4}
\end{figure}
\begin{figure}
 \vspace*{-0.5cm}
\begin{center}
  \includegraphics[width=320pt, height=240pt, angle=0]{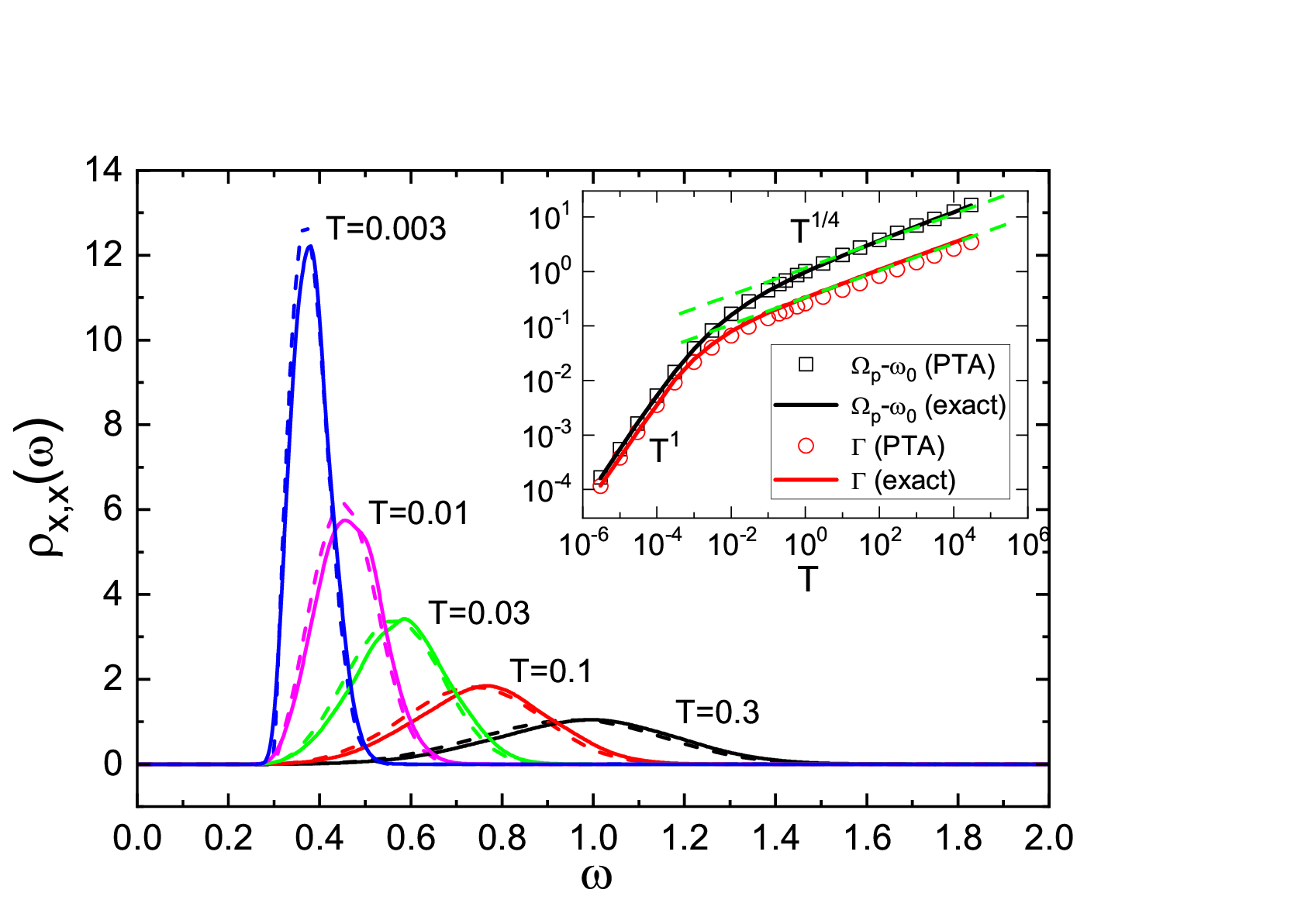}
  \vspace*{-1.0cm}
\end{center}
  \caption{(color online) Spectral function $\rho_{x,x}(\omega)$ of AHO model obtained by averaging the results of $500$ random realization of $\{ \lambda_i \}$ on basis $B_2 =\{x e^{\lambda_i H} \}$ ($i=1, 2, ..., N$) (solid lines). The broadening scheme $\sigma = r \Gamma$ is used with $r=0.3$. The exact curve is dashed line. Inset: relative peak position $\Omega_p - \omega_0$ (squares) and the peak width $\Gamma$ (circles) as functions of $T$ from PTA. The solid lines are extracted from exact spectral function. Dashed lines mark the power law $T^{1/4}$ behavior. Basis dimension $N=20$. Model parameters are $\mu=1.0$, $\omega_0 = 0.3$, and $\alpha = 0.25$. $\Delta = 6.0$ is used.
}   \label{Fig5}
\end{figure}
Fig.4 shows the spectral function $\rho_{x,x}(\omega)$ obtained from PTA on the $H$-expanded basis $B_2$ (red solid lines), together with the exact curve (black dashed lines). They are shown in log-linear axes. In contrast to the case of basis $B_1$ (shown in Fig.3), here as the dimension of basis increases, in the frequency regime of the primary peak, one gets significantly more delta peaks. The envelope tends to the exact broad primary peak much faster than that of $B_1$ basis. A basis with dimension $N=20$ already produces about $10$ visible peaks densely located within the frequency regime of primary peak, leading to qualitatively accurate description of the primary peak, if each delta peak of PTA is properly broadened with a Gaussian function. However, $B_2$ basis also meets its limitation at about $N=30$, beyond which the spectral function can hardly receive any improvement. This is due to the deterioration of the condition number of ${\bf I}$ matrix as the basis dimension increases.

In Fig.5, we compare the spectral function $\rho_{x,x}(\omega)$ averaged over $500$ random $\lambda_i$ realizations (solid lines) with the exact curve (dashed lines) for different temperatures. Only the primary peak is shown in this plot with a linear-linear axis. The agreement is quite good in a wide temperature range. The exact $\rho_{x,x}(\omega)$ curve has a natural frequency cut off at $\omega = \omega_0$, below which there is no spectral weight. This feature is nicely captured by PTA on basis $B_2$ with averaging on $500$ random $\{ \lambda_i \}$'s (not shown in Fig.5). The inset shows that the relative peak position $\Omega_p - \omega_0$ and the peak width $\Gamma$ from Eqs.(\ref{Eq48}) and (\ref{Eq49}) are in qualitative agreement with the exact one from low- to high-temperature limits. The quantitative deviation in $\Gamma$ becomes apparent only in the high $T$ regime, with PTA value lower by a factor $0.77$. $\Omega_p - \omega_0$ and $\Gamma$ have the same power law of $T$, being $T^{1}$ and $T^{1/4}$ in the low and high temperature limits, respectively. 

These power laws can be understood as follows. The peak position $\Omega_p$ is qualitatively described by the $N=1$ PTA expression
\begin{equation}   \label{Eq50}
  \omega_p = \sqrt{\omega_0 + (12 \alpha/ \mu) \langle x^2 \rangle}.
\end{equation}
In the low temperature limit, the system tends to be harmonic, $\langle x^2 \rangle \sim T \ll \omega_0$ due to equipartition theorem. In the high temperature limit, the particle's movement is dominated by the $x^{4}$ potential. Exact analysis gives $\langle x^2 \rangle \sim T^{1/2} \gg \omega_0$ \cite{Ma2}. Equation (\ref{Eq50}) then gives the asymptotic behavior of $\Omega_p(T)$ observed in the inset of Fig.5,
\begin{equation}
 \Omega_{p}(T) \sim 
   \left\{\begin{array}{lll}
       \omega_0 + c \, T^{1}, & \,\,\,  (T  \ll T_{cr}) \\
        \\
       c^{\prime} \, T^{\frac{1}{4}},      & \,\,\,  (T \gg T_{cr}).
     \end{array} \right.   
\end{equation}
Here, $c$ and $c^{\prime}$ are $T$-independent coefficients. The crossover temperature $T_{cr} \approx 1.0$.

The spectral width $\Gamma(T)$ is generated by the thermal fluctuation of oscillating frequency which, due to the nonlinear potential, comes from the thermal fluctuation of the initial energy $E_0$ of the system. Guided by Eq.(\ref{Eq50}), we assume that the particle oscillates with a frequency $\omega \sim \sqrt{\omega_0^{2} + 12 \alpha x_0^{2}} $, with $x_0$ being the initial coordinate. $x_0$ depends on $E_0$ through $E_0 \sim \frac{1}{2}\mu \omega_0 x_0^{2}$ at low $T$, and $E_0 \sim \alpha x_0^{4}$ at high $T$. Inserting them into the expression for $\omega$ and considering $E_0 \sim \delta E_0 \sim T$ due to thermal fluctuation, we obtain the fluctuation of frequency $\delta \omega \sim \delta E_0 \sim T$ at low $T$, and $\delta \omega \sim E_0^{-3/4}\delta E_0 \sim T^{1/4}$. This  explains the observed asymptotic behavior of $\Gamma(T)$ in the inset of Fig.5,
\begin{equation}
 \Gamma(T) \sim 
   \left\{\begin{array}{lll}
       d \, T^{1}, & \,\,\,  (T  \ll T_{cr}) \\
        \\
       d^{\prime} \, T^{\frac{1}{4}},      & \,\,\, (T \gg T_{cr}).
     \end{array} \right.   
\end{equation}
\begin{figure}
 \vspace*{-0.0cm}
\begin{center}
  \includegraphics[width=280pt, height=210pt, angle=0]{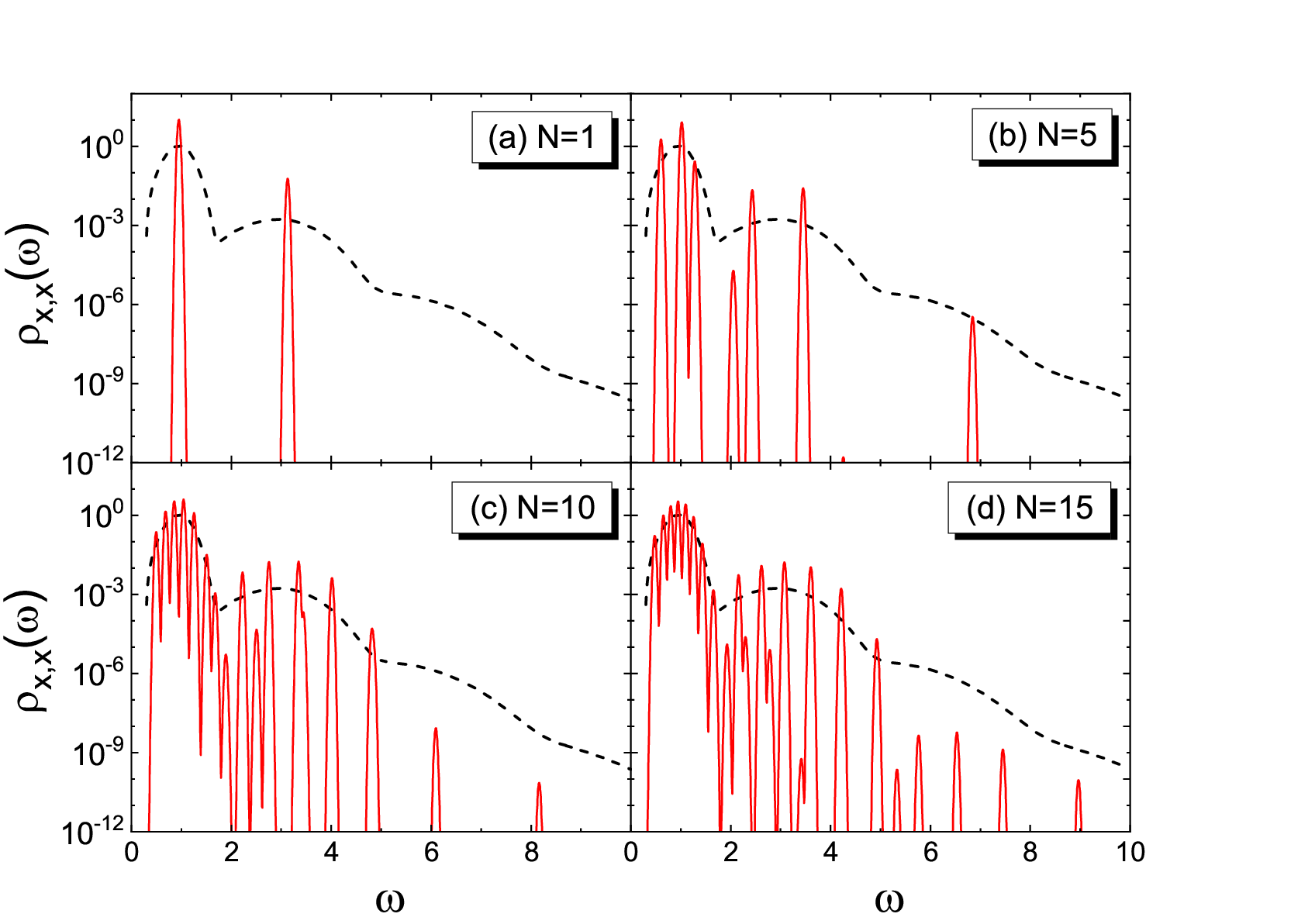}
  \vspace*{-1.0cm}
\end{center}
  \caption{(color online) Spectral function $\rho_{x,x}(\omega)$ of AHO model from basis $B_3 = \{x, x^{3}\} \otimes \{ e^{\lambda_i H} \}$ ($i=1,2,...,N$) (red solid line), broadened with $\sigma=0.02$. The exact curve is the black dashed line. Model parameters are $\mu=1.0$, $\omega_0 = 0.3$, $\alpha = 0.25$, and $T=0.3$. $\Delta = 6.0$ is used.
}   \label{Fig6}
\end{figure}

If we further enlarge the basis $B_2$ by adding $\{ x^3 e^{\lambda_i H} \}$, we obtain basis $B_3$. Fig.6 shows the spectral function $\rho_{x,x}(\omega)$ from PTA on basis $B_3$. The exact curve is shown as the black dashed line. The basis dimension of $B_3$ is $D=2N$. At $N=1$, the two delta peaks are located at the central positions of the primary and the secondary peaks of the exact curve. With increasing $N$, PTA produces more and more delta peaks in the frequency regime of the primary and the secondary peaks, forming two broadened curves. Combining the results for $B_2$ and $B_3$ bases, we conclude that the $H$-expanding of basis can lift the degeneracy of the quasi-particle excitation energies obtained from a lower order PTA, and split the corresponding delta peak into a bunch of densely located peaks, which effectively describe a thermally broadened peak.

\end{subsection}
%---------------------------------------------------%

\end{section}

%========================================================================%%
\begin{section}{$\mathcal{H}_k$-expanded basis for the 1D nonlinear $\phi^4$ lattice model}

In this section, we apply the $H$-expanded basis formalism to the classical 1D nonlinear $\phi^4$ lattice model, and we calculate the phonon spectral function $\rho_{Q_k, Q_k^{\ast}}$, with a focus on the thermal broadening effect of the phonon peak at $T>0$. Here, $Q_{k} = 1/\sqrt{L} \sum_{j} e^{-ijk}x_j$ is the Fourier component of coordinates of particles. The Hamiltonian of the $\phi^4$ model reads
\begin{equation}
  H = \sum_{i=1}^L \Big[ \frac{p_i^2}{2m} + \frac{K}{2} (x_i-x_{i-1})^2 + \frac{\gamma}{4} x_i^4 \Big],
\end{equation}
Here, $L$ is the total number of particles, $x_i$ represents the deviation of the $i$-th particle from its equilibrium position, $K$ is the nearest-neighbor coupling strength, and $\gamma$ is the coefficient of the on-site potential. We use periodic boundary condition $x_{0}=x_{L}$. This model is obtained by discretizing the classical $\phi^4$ field \cite{Boyanovsky1}. It has been widely used in the study of 1D heat transport \cite{Aoki1, BH33, KA34, AD35, Li3, Xu1, Xu2, NL38}. Similar to the FPU-$\beta$ model \cite{KA39}, the $\phi^4$ model has a scaling property which makes it sufficient to study only the parameter point $K = \gamma = 1$ \cite{Ma2}. Below, we will focus on this parameter point.

For this model, the previous PTA study employs the basis $Q_{k}$ \cite{Ma2}. To describe the thermal fluctuation effect, it is tempting to expand it into the basis $A^{k}_{i} \equiv Q_{k} e^{\lambda_{i} H}$ ($i=1,2,...,N$), similar to the case of one-variable AHO model. However, we find that this scheme has two problems. First, the peak width $\Gamma_k$ in the phonon dispersion function $\rho_{Q_k, Q_k^{\ast}}(\omega)$ obtained from this basis is proportional to $1/\sqrt{L}$, which vanishes in the large size limit $L \rightarrow \infty$. This is unphysical. The reason is that, unlike in the one-variable AHO model, the thermal broadening of the phonon peak of $\rho_{Q_k, Q_k^{\ast}}(\omega)$ for a long chain of atoms is no longer contributed by the fluctuation of total energy $H$, but mainly by the thermal fluctuation of energy $\mathcal{H}_k$ of mode $k$. Indeed, in the microscopic canonical ensemble where the total energy $H$ is fixed, the $H$-expanded basis becomes identical to the single variable $Q_k$ and hence cannot produce a broadened phonon peak, while the MD simulation in the microscopic canonical ensemble can correctly produce the phonon broadened effect.

In fact, in the canonical ensemble in which our formalism is written, the thermal fluctuation $\delta H$ of the total energy $H$, when distributed equally to each mode, gives a magnitude $\delta H / L$ to the fluctuations of each mode $H_k$. Due to the nonlinear coupling, it further gives an uncertainty of the frequency $\delta \omega_k \sim \sqrt{ \delta H/L}$. This explains the $1/\sqrt{L}$ dependence of the incorrect results obtained from the basis $Q_{k} e^{\lambda_{i} H}$. 

The second problem of the basis $\{Q_{k} e^{\lambda_{i} H} \}$ is that the complex variable $Q_k$ becomes real at momenta $k=0$ and $k=\pi$, which have distinct dynamical and statistical properties from the complex variables at $k \neq 0, \pi$. As a result, special care must be taken in PTA formalism to avoid the possible discontinuity in physical results at $k=0$ and $\pi$. This will complicate the formalism. 

Due to the considerations above, in the following, we will work on the real basis $\text{Re} Q_k$ and $\text{Im} Q_k$, and we expand the basis using the effective local energy $\mathcal{H}_k$ of the mode $k$, instead of the total Hamiltonian $H$. Although $\mathcal{H}_k$ is no longer a strictly conserving quantity like $H$, it is approximately conserving on the level of $Q_k$-PTA.

\begin{subsection}{ $\{ Q_k\}$-PTA reformulated on a real basis}

Since we will construct the $\mathcal{H}_k$-expanded real basis, in this subsection, we first derive the PTA formalism on the real basis $\{ \text{Re}{Q_k}, \text{Im}{Q_k} \}$. The obtained equations are equivalent to those of PTA on the complex basis $\{ Q_k \}$ discussed in Ref.\cite{Ma2}. Then we expand it with $e^{\lambda_i \mathcal{H}_{k}}$, where $\mathcal{H}_{k}$ is the effective energy of the real mode $k$ (see below). In this part, we reformulate this PTA in terms of the real basis, because we will need the results of this lower order PTA to identify the approximate eigen-exciatation variable in order to introduce the proper mode energy $\mathcal{H}_{k}$ and to expand the former with the latter.

We first define the following dynamical variables,
\begin{eqnarray}
   && Q_k \equiv \frac{1}{\sqrt{L}} \sum_j x_j e^{-ijk},  \nonumber\\
   && P_k \equiv \frac{1}{\sqrt{L}} \sum_j p_j e^{ijk},  \nonumber\\
   && R_k \equiv \frac{1}{\sqrt{L}} \sum_j x^3_j e^{-ijk}. 
\end{eqnarray}
The lattice momentum $k$ takes $L$ different values $k = (2\pi/L) n_k$ ($n_k = 0, 1, ..., L-1$).
Here and below, we assume that $L$ is an even number so that $k=\pi$ is possible. We take the real and imaginary part of $Q_k$ and $P_k$ as our dynamical variables. Considering the independence of variables, we will confine the momentum $k$ to $[0, \pi]$, and we take the following definition for the dynamical variables
\begin{eqnarray}
  && Q_{1k} \equiv \alpha_k \text{Re}{Q_k},   \,\,\, \,\,\,\, k \in [0, \pi], \nonumber \\
  && Q_{2k} \equiv -\alpha_k \text{Im}{Q_k},    \,\,\,\,\,\,\, k \in (0, \pi), \nonumber\\ 
  && P_{1k} \equiv \alpha_k \text{Re}{P_k},   \,\,\, \,\,\,\, k \in [0, \pi], \nonumber \\
  && P_{2k} \equiv \alpha_k \text{Im}{P_k},    \,\,\,\,\,\,\, k \in (0, \pi).
\end{eqnarray}
In the above equations, the factor $\alpha_k$ is for normalizing the variables. $\alpha_k = 1$ for $k=0$ and $\pi$, and $\alpha_k = \sqrt{2}$ for $k \neq 0$ or $\pi$.
Note that since $Q_{k=0}$ and $Q_{k=\pi}$ are real variables, for $Q_{1 k}$ and $P_{1 k}$, $k$ takes $L/2+1$ values, $k = (2\pi/L) n_{k}$ ($n_k = 0, 1, ..., L/2$). For $Q_{2 k}$ and $P_{2 k}$, $k$ takes $L/2-1$ values, $k = (2\pi/L) n_{k}$ ($n_k = 1, ..., L/2-1$). Altogether we have $L$ independent real variables.

One can prove that the above-defined variables $Q_{\nu k}$ and $P_{\nu k}$ ($\nu=1,2$) fulfill the properties of canonical variables. That is, $\{Q_{\nu k}, Q_{\mu p} \} = \{P_{\nu k}, P_{\mu p} \} = 0$ and $\{Q_{\nu k}, P_{\mu p} \} = \delta_{\nu \mu} \delta_{kp}$. They are a set of complete and independent real dynamical variables in place of $\{x_i, p_i \}$ or $\{Q_k, P_k \}$. 

The PTA formalism on the basis $Q_{\nu k}$ ($k \in [0, \pi]$ for $\nu = 1$, and $k \in (0, \pi)$ for $\nu=2$) can be obtained from the previous PTA equations on basis $Q_{k}$ \cite{Ma2}. We obtain the inner product matrix $I_{\nu k, \mu p}$ and the Liouville matrix $L_{\nu k, \mu p}$ as
\begin{eqnarray}    \label{Eq56}
  && I_{\nu k, \mu p} = \frac{1}{m} \delta_{\nu \mu} \delta_{kp},  \nonumber \\
  && L _{\nu k, \mu p} = \frac{1}{m} \omega_{k}^{2} \delta_{\nu \mu} \delta_{kp}.
\end{eqnarray}
The phonon excitation energy $\omega_k$ is given by 
\begin{equation}    \label{Eq57}
  \omega_k = \sqrt{ \frac{1}{m} \left[2K\left( 1- \cos{k} \right) + 3 \gamma \langle x^{2} \rangle \right] }. 
\end{equation}
From these expressions, one obtains the spectral functions $\rho_{Q_{\nu k}, Q_{\mu p} }(\omega)$ as
\begin{equation}
   \rho_{Q_{\nu k}, Q_{\mu p} }(\omega) = \delta_{\nu \mu} \delta_{kp} \frac{1}{2m \omega_k} \left[\delta(\omega - \omega_k ) - \delta(\omega + \omega_k) \right].
\end{equation}

The PTA self-consistent equation for $\langle x^2 \rangle$ is obtained from the spectral theorem Eq.(\ref{Eq4}) as
\begin{eqnarray}     \label{Eq59}
  \langle x^2 \rangle &=& \frac{1}{L} \sum_{\nu=1}^{2} \sum_{k_{\nu}} \langle Q_{\nu k_{\nu}}^{2} \rangle   \nonumber \\
  &=& \frac{1}{L} \sum_{k \in [0, 2\pi)} \frac{1}{\beta m \omega_{k}^{2}}.
\end{eqnarray}
In the last line above, we have used the fact that $\langle Q_{1k}^{2}\rangle =\langle Q_{2k}^{2}\rangle = \langle Q_{1 \, 2\pi-k}^{2}\rangle $ for $k \in (0, \pi)$. The final expression is same as that of the $\{ Q_k \}$ basis PTA.

\end{subsection}

\begin{subsection}{ effective Hamiltonian $\mathcal{H}_{\nu k}$ of the mode $(\nu k)$}

In order to expand the basis $Q_{\nu k}$ using the mode energy, we first have to construct an effective energy $\mathcal{H}_{\nu k}$ for mode $(\nu k)$. We will do this construction approximately based on the real basis $\{Q_{1k}, Q_{2k} \}$ PTA equations discussed above. 
The effective energy $\mathcal{H}_{\nu k}$ is constructed by the following principles. First, it is a real variable with the unit of energy; second, it generates the expected dynamics, {\it i.e.}, $\{ Q_{\nu k}, \mathcal{H}_{\nu k} \} \approx \{ Q_{\nu k}, H \}$ and $\{ P_{\nu k}, \mathcal{H}_{\nu k} \} \approx \{ P_{\nu k}, H \}$; and third, it is a zero-th order conserving quantity, {\it i.e.}, $\{\mathcal{H}_{\nu k}, H \} \approx 0$. The third principle is required because the broadening of the phonon peak will be regarded in our theory as splitting the degeneracy of the excitation energy $\omega_k$ by high order fluctuations beyond the zero-th order approximation, as discussed in the case of the one-variable AHO model. In this work, the two last principles are fulfilled at the level of PTA on the basis $\{Q_{1k}, Q_{2k} \}$.

From Eq.(\ref{Eq56}) of $\{Q_{1k}, Q_{2k} \}$-PTA and the general principle of PTA, we have the approximate EOM as
\begin{eqnarray}
   && \{ Q_{\nu k}, H \} = \frac{1}{m} P_{\nu k},   \nonumber \\
   && \{ P_{\nu k}, H \} \approx -m \omega_{k}^{2} \, Q_{\nu k}.   \,\,\,\,\,\,\, (\nu = 1,2)
\end{eqnarray}
It is then easy to find the two eigen excitation variables associated to mode $(\nu k)$ by diagonalizing the EOM matrix. We obtain
\begin{eqnarray}
   && a_{\nu k}^{(1)} = c_k \left(m \omega_{k}^{2} \, Q_{\nu k} - i\omega_k P_{\nu k} \right),  \nonumber \\
   && a_{\nu k}^{(2)} = \tilde{c}_k \left(m \omega_{k}^{2} \, Q_{\nu k} + i\omega_k P_{\nu k} \right).
\end{eqnarray}
They obey $\{a_{\nu k}^{(1)}, H \} \approx i \omega_k \, a_{\nu k}^{(1)}$, and  $\{a_{\nu k}^{(2)}, H \} \approx -i \omega_k \, a_{\nu k}^{(1)}$, respectively.

One can construct the approximately conserving quantity $\mathcal{H}_{\nu k}$ as $\mathcal{H}_{\nu k} = f_{k} a_{\nu k}^{(1) \ast} a_{\nu k}^{(1)} $. It satisfies the approximate conservation $\{ \mathcal{H}_{\nu k}, H \} \approx 0$. Fixing the coefficients $c_{k}$ and $\tilde{c}_k$ from the second principle, {\it i.e.}, $\{ Q_{\nu k}, \mathcal{H}_{\nu k}  \} \approx \{ Q_{\nu k}, H \}$, we obtain the effective Hamiltonian of mode $(\nu k)$ as 
\begin{equation}
   \mathcal{H}_{\nu k} =  \frac{1}{2} m \omega_{k}^{2} \, Q_{\nu k}^{2} + \frac{1}{2m} P_{\nu k}^{2},
\end{equation}
which has the desired properties
\begin{eqnarray}   \label{Hk}
  &&   \{ \mathcal{H}_{\nu k}, H \} \approx 0, \nonumber \\
  && \{ Q_{\nu k}, \mathcal{H}_{\nu k} \} \approx \{ Q_{\nu k}, H \}   \nonumber \\
  && \{ P_{\nu k}, \mathcal{H}_{\nu k} \} \approx \{ P_{\nu k}, H \}.
\end{eqnarray}

\end{subsection}

\begin{subsection}{ PTA on $\mathcal{H}_{\nu k}$-expanded basis}

Following the same practice of the PTA on the $H$-expanded basis for the one-variable AHO model, we construct the expanded basis for the 1D $\phi^4$ lattice model in the subspace $(\nu k)$ as
\begin{equation}    \label{Eq62}
   A^{\nu k}_{i} = Q_{\nu k} e^{\lambda_i \mathcal{H}_{\nu k}}.  \,\,\,\,\, (\lambda_i \,\,\, \text{real},\,\,\,\, i =1, 2, ..., N)
\end{equation}
Here, $(\nu k) = (1, k_1)$ and $(2, k_2)$, with $k_{1} \in [0, \pi]$ and $k_2 \in (0, \pi)$, respectively. We assign $\lambda_1 = 0$ to make sure $A^{\nu k}_1 = Q_{\nu k}$. 

The inner product matrix ${\bf I}^{\nu k}$ and the Liouville matrix ${\bf L}^{\nu k}$ are derived from straightforward
calculation, along the line for the AHO model. In the derivation, we have used the properties Eq.(\ref{Hk}) of $\mathcal{H}_{\nu k}$ . We obtain
\begin{eqnarray}    \label{Eq63}
   I^{\nu k}_{ij} &=& \frac{1}{m} \frac{\beta}{\beta - \lambda_i - \lambda_j} \langle e^{(\lambda_i + \lambda_j) \mathcal{H}_{\nu k} } \rangle,  \nonumber \\
   L^{\nu k}_{ij}  &=&  \frac{1}{m} 2K(1-\cos{\omega}) I^{\nu k}_{ij}  \nonumber \\
        && + \frac{1}{m^2} \frac{\beta}{\beta - \lambda_i - \lambda_j}  3 \gamma \langle x^2 e^{(\lambda_i + \lambda_j) \mathcal{H}_{\nu k} } \rangle.   
\end{eqnarray}

In the above equations, like in the study of AHO model, we evaluate $\langle e^{(\lambda_i + \lambda_j) \mathcal{H}_{\nu k} } \rangle$ using the quadratic variational approximation
\begin{eqnarray}    \label{Eq64}
  \langle e^{(\lambda_i + \lambda_j) \mathcal{H}_{\nu k} } \rangle &\approx& \langle e^{(\lambda_i + \lambda_j) \mathcal{H}_{\nu k} } \rangle_{vari} \nonumber \\
  & = & \frac{\beta}{\beta - \lambda_i - \lambda_j},
\end{eqnarray}
with the variational Hamiltonian 
\begin{eqnarray}
 &&  H_{vari} = \sum_{k} H_k,  \nonumber \\
 &&  H_{k} = \frac{1}{2m} P_{k}^{\ast}P_{k} + \frac{1}{2} m \omega_{k}^{2} Q_{k}^{\ast}Q_{k}.
\end{eqnarray}

To calculate $\langle x^2 e^{(\lambda_i + \lambda_j) \mathcal{H}_{\nu k} } \rangle$ self-consistently, we write it as
\begin{eqnarray}     \label{Eq69}
&& \langle x^2 e^{(\lambda_i + \lambda_j) \mathcal{H}_{\nu k} } \rangle   \nonumber \\
 &=& \frac{1}{L} \sum_{\mu p} \langle Q_{\mu p}^{2} e^{(\lambda_i + \lambda_j) \mathcal{H}_{\nu k} } \rangle \nonumber \\
 &=& \frac{1}{L} \sum_{\mu p} \langle Q_{\mu p}^{2} e^{(\lambda_i + \lambda_j) \mathcal{H}_{\mu p} } \frac{ e^{(\lambda_i + \lambda_j) \mathcal{H}_{\nu k} }} { e^{(\lambda_i + \lambda_j) \mathcal{H}_{\mu p} }} \rangle \nonumber \\
 & \approx & \frac{1}{L} \sum_{\mu p} \left[ \langle Q_{\mu p}^{2} e^{(\lambda_i + \lambda_j) \mathcal{H}_{\mu p} } \rangle \frac{ \langle  e^{(\lambda_i + \lambda_j) \mathcal{H}_{\nu k} } \rangle }{ \langle  e^{(\lambda_i + \lambda_j) \mathcal{H}_{\mu p} } \rangle} \right]   \nonumber \\
 & \approx & \frac{1}{L} \sum_{\mu p} \langle A^{\mu p}_{i} A^{\mu p}_{j} \rangle .
\end{eqnarray}
In the above approximation, the momentum dependence of $\langle x^2 e^{(\lambda_i + \lambda_j) \mathcal{H}_{\nu k}} \rangle$ is neglected. This makes the interaction part of the $\bf{L}$ matrix $k$-independent, reminiscent of the local self-energy in the dynamical mean-field theory \cite{Vollhardt1,Kotliar1}. Note that the additional approximation used in Eq.(\ref{Eq69}) exaggerates the coupling between different modes and tends to overestimate the broadening. See Fig.{\ref{Fig11}}. 
The final expression can be evaluated from ${\bf I}^{\nu k}$ and ${\bf L}^{\nu k}$ via
\begin{equation}    \label{Eq70}
   \langle A^{\nu k}_{i} A^{\nu k}_{j} \rangle \approx \frac{1}{\beta} \left[{\bf I}^{\nu k} \left( {\bf L}^{\nu k} \right)^{-1} {\bf I}^{\nu k}  \right]_{ij}.
\end{equation}

From the self-consistent solution of the above equations, one can finally obtain the physical quantities of interest, {\i.e.}, the spectral function $\rho_{Q_{k}, Q_{k}^{\ast}}(\omega)$. It is expressed by the spectral functions of real variables $\rho_{Q_{1k}, Q_{1k}}(\omega)$ as
\begin{equation}
 \rho_{Q_{k}, Q_{k}^{\ast}}(\omega) =
   \left\{\begin{array}{lll}
        \rho_{Q_{1k}, Q_{1k}}(\omega), & \,\,\, k \in [0, \pi] \\
        \\ 
\rho_{Q_{1 \, 2\pi-k}, Q_{1 \, 2\pi-k}}(\omega), & \,\,\, k \in (\pi, 2\pi). 
     \end{array} \right.   
\end{equation}
The spectral function $\rho_{Q_{1k}, Q_{1k}}(\omega)$ is then calculated from the standard PTA formalism Eqs.(\ref{Eq11}) and (\ref{Eq12}). Note that in the case of $N=1$, the $\mathcal{H}_{\nu k}$-expanded basis recovers the basis $\{Q_{\nu k}\}$, and the excitation energy is given by the $\omega_k$ in Eq.(\ref{Eq57}).

In the practical calculation, the above equations are combined with the zeros-removing method described in section III.C.

\end{subsection}
%-------------------------------------

\begin{subsection}{spectral function $\rho_{Q_k, Q_k^{\ast}}(\omega)$ for 1D $\phi^4$ lattice model}

\begin{figure}
 \vspace*{-2.0cm}
\begin{center}
  \includegraphics[width=380pt, height=290pt, angle=0]{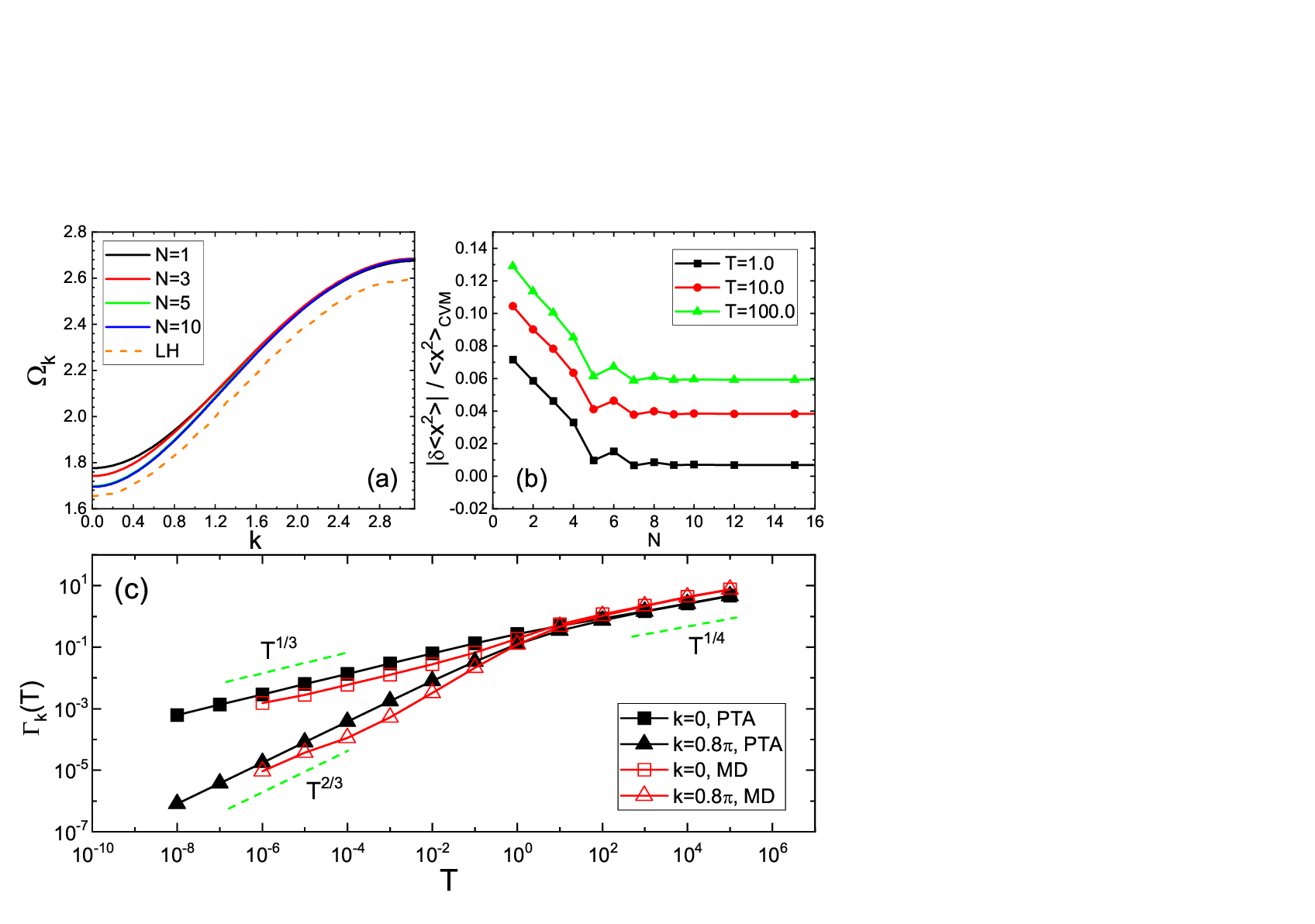}
  \vspace*{-1.0cm}
\end{center}
  \caption{(color online) (a) Phonon dispersion $\Omega_k$ curves at $T=5.0$ from PTA on the $\mathcal{H}_{\nu k}$-expanded basis, for various dimension $N$'s (solid lines). The dashed line (denoted as LH) is the lower bound harmonic variational result \cite{Liu1}. (b) Relative error of $\langle x^2 \rangle$ from PTA with respect to that from cluster variational method (CVM) \cite{Jia1} versus basis dimension $N$, for different temperatures. (c) Spectral width $\Gamma_k$ as functions of temperature. Filled and empty symbols with guiding lines are data from PTA and MD, respectively. The dashed lines mark the power law. Model parameters are $m=1.0$, $K=1.0$, $\gamma=1.0$, and $L=1000$. $L=4000$ is used for $T \leq 10^{-2}$ in panel (c).
}   \label{Fig7}
\end{figure}
\begin{figure}
 \vspace*{-2.5cm}
\begin{center}
  \includegraphics[width=400pt, height=320pt, angle=0]{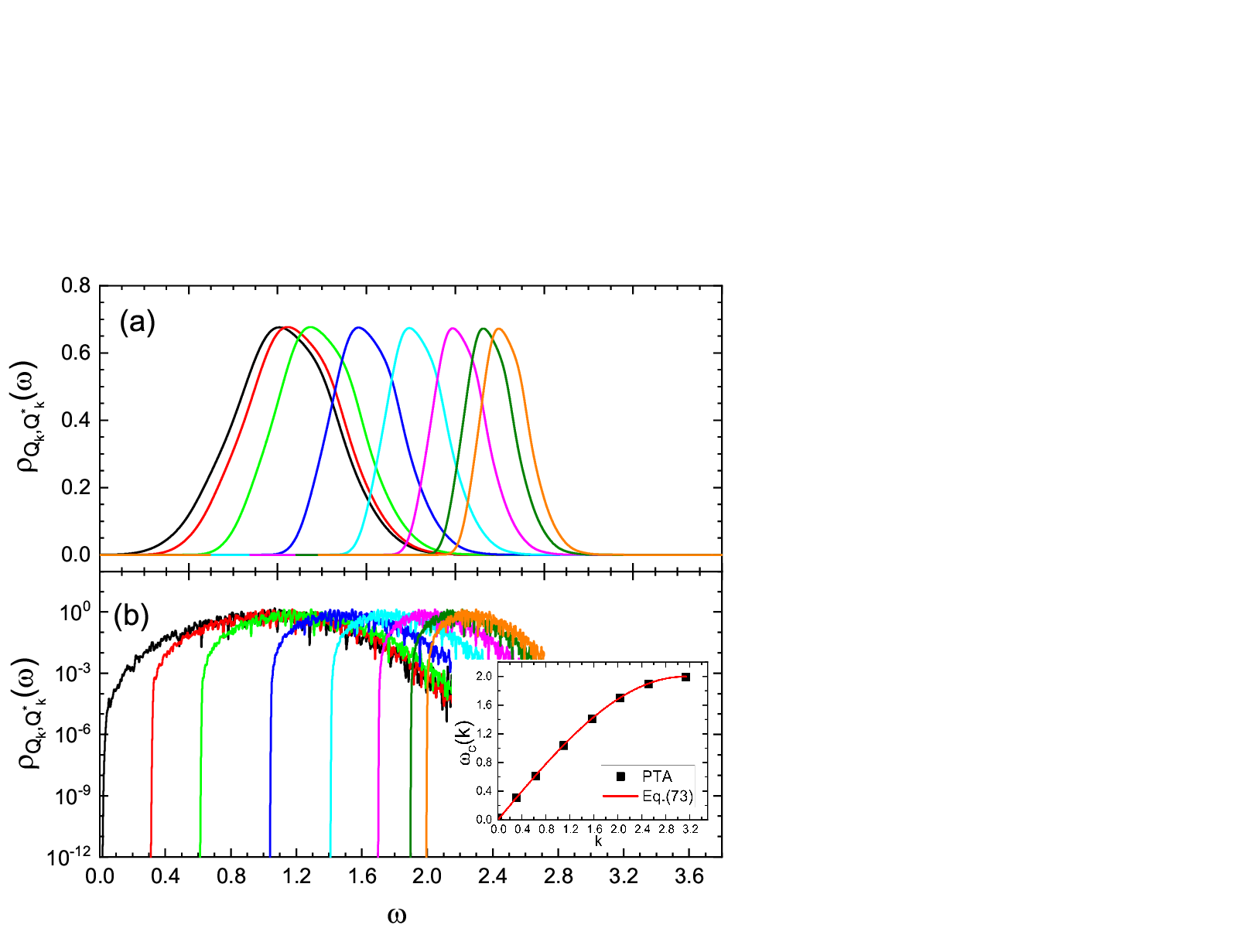}
  \vspace*{-1.0cm}
\end{center}
  \caption{(color online) Spectral function $\rho_{Q_k, Q_k^{\ast}}(\omega)$ 
  at various $k$ for 1D $\phi^4$ lattice model at $T=1.0$. In both panels, from left to right, $k=0$, $0.1\pi$, $0.2\pi$, $0.35\pi$, $0.5\pi$, $0.65\pi$, $0.8\pi$, and $\pi$. (a) linear scale, with broadening parameter $r=0.3$; (b) log scale, with $r=0.01$, obtained by averaging results of $200$ random $\{\lambda_i\}$'s. Inset of panel (b) shows the cut-off frequency $\omega_c(k)$ extracted from PTA (squares) and Eq.(\ref{cutoff}) (line). Model parameters are $m=1.0$, $K=1.0$, $\gamma=1.0$, and $L=1000$. $\Delta=7.0$ for $k \neq 0$ and $\Delta=12.0$ for $k=0$.
}   \label{Fig8}
\end{figure}

Here, we present the numerical results obtained from the above formula for 1D $\phi^4$ lattice model. We use the scheme Eqs.(\ref{Eq46}) and (\ref{Eq47}) for assigning $\{ \lambda_i \}$ values in our calculation.

First, in Fig.\ref{Fig7}, we benchmark PTA on the $\mathcal{H}_{\nu k}$-expanded basis by checking the convergence of its results with increasing basis dimension $N$, taking other results as references. In Fig.\ref{Fig7}(a), we plot the dispersion $\Omega_k$ curve obtained from successively larger bases. The curve from the lower bound harmonic variational approximation (LH) \cite{Liu1} (dashed line) is shown for comparison. It is seen that as $N$ increases, the long-wavelength part of $\Omega_k$ shifts towards the LH curve which is expected to be more accurate, but the short-wavelength part shifts only slightly. The $\Omega_k$ curve converges at about $N=10$. Since our basis is not complete in the limit $N=\infty$, we do not expect the dispersion to converge to the exact curve. We see that the PTA with $N=1$ already produces quite accurate $\Omega_k$ and for this quantity only limited improvement can be gained by enlarging $N$.

Fig.\ref{Fig7}(b) shows the relative error in $\langle x^2 \rangle$ calculated from PTA, taking the value from a two-site cluster variational method (CVM) \cite{Kikuchi1,Lopez1} as a reference. The CVM was originally designed for Ising-like lattice models with discrete degrees of freedom and was applied to the Klein-Gordan lattice model with continuous degrees of freedom \cite{Jia1}.  It is known that the two-site CVM produces exact static averages for 1D models with nearest-neighbour interaction \cite{Pelizzola1}. So here we take the CVM results as a reference. As $N$ increases from $N=1$, the error of PTA first decreases and then saturates at about $N=10$. The saturated error increases with increasing temperature. This shows both the effect and the limitation of the $\mathcal{H}_{\nu k}$-expanded basis.

In Fig.\ref{Fig7}(c), the spectral widths $\Gamma_{k=0}(T)$ and $\Gamma_{k=0.8\pi}(T)$ extracted from $\rho_{Q_k, Q_k^{\ast}}(\omega)$ of PTA are compared with those from MD simulations. In MD simulations, $\Gamma_k(T)$ is extracted from the power spectrum using the half-area method \cite{Guo1}. For this purpose, we first define the complex normal variables
\begin{align} \label{eq:ak}
  a_k = i \tilde{P}_k + \tilde{Q}_k,
\end{align}
where 
\begin{align}
  \tilde{P}_k &= \sqrt{\frac{2}{L}} \, \sum_{j = 1}^{L} p_j \sin(jk),\\
  \tilde{Q}_k &= \sqrt{\frac{2}{L}} \, \sum_{j = 1}^{L} x_j \sin(jk).
\end{align}
are momentum and coordinate variables. We calculate
$\vert \hat{a}_k(\omega)\vert^2$, the Fourier transform of the complex normal variable $a_k(t)$ in Eq.~(\ref{eq:ak}). $\Gamma_k$ is then defined as the width of the frequency window around the center $\tilde{\omega}_k$ of the power spectrum $\vert \hat{a}_k(\omega)\vert^2$, in which the power is equal to half of the total power. That is, 
\begin{align}
  \frac{\displaystyle\int_{\tilde{\omega}_k-\Gamma_k/2}^{\tilde{\omega}_k + \Gamma_k/2} \vert \hat{a}_k(\omega)\vert^2 \, d\omega}{\displaystyle\int_{0}^{\infty}\vert \hat{a}_k(\omega)\vert^2 \, d\omega} = \frac{1}{2}.
\end{align}
Here, $\tilde{\omega}_k$ is the center frequency of the peak in $\vert \hat{a}_k(\omega)\vert^2$.
In the case of large fluctuations, this definition gives a much more robust result in the numerical simulation than the standard method does \cite{Guo1}.

In Fig.\ref{Fig7}(c), for both $k=0$ and $k=0.8\pi$, we see a qualitative agreement between PTA and MD results. 
Both curves for $k=0$ (squares) and $k=0.8\pi$ (up triangles) show qualitative agreement in all temperature regime.
In particular, the asymptotic power law behaviors, {\it i.e.}, $\Omega_k(T \rightarrow \infty) \sim T^{1/4}$, $\Omega_{k=0}(T \rightarrow 0) \sim T^{1/3}$, and $\Omega_{k \neq 0}(T \rightarrow 0) \sim T^{2/3}$ are observed in both results. The source of these asymptotic power laws will be analysed in Fig.\ref{Fig12}. Part of the quantitative deviation between PTA and MD results may be traced back to the truncation error and additional errors introduced in Eq.(\ref{Eq69}). We find that the present PTA on the $H_{\nu k}$-expanded basis slightly underestimates $\Gamma_k$ in the high-temperature regime, and overestimate it by a factor of $5.0$ in the low-$T$ regime. This is similar to the case of the AHO model, as shown in the inset of Fig.5. Another source of the discrepancy may lie in the different definition and extraction method of $\Gamma_k$. Other spectral width data from MD simulation in the literatures \cite{AP1,Xu2,YLiu1} are also compared with our results. We find qualitative agreement in both the $k$ and $T$ dependences. See Fig.\ref{Fig11}.

In Fig.\ref{Fig8}, we show the evolution of spectral function $\rho_{Q_k, Q_k^{\ast}}(\omega)$ of 1D $\phi^4$ lattice model with momentum $k$ at $T=1.0$, obtained from $\mathcal{H}_{\nu k}$-expanded basis PTA and averaging over $200$ random realization of $\{ \lambda_i \}$'s. Panels (a) and (b) show the curve on the linear and log y axis, respectively. As $k$ increases, the peak position increases and the width of the peak shrinks. The heights of the peaks do not change much, since the total weight of $\rho_{Q_k, Q_k^{\ast}}(\omega)$ in $\omega > 0$ is not conserved. In (b), the spectral function shows a feature similar to that of AHO model: a low frequency cut-off at frequency $\omega_c(k)$. In AHO model, $\rho_{x,x}(\omega)$ has a low frequency cut-off at $\omega_{c} = \omega_0$, which is well described by the $H$-expanded PTA with random averaging. The similarity between PTA equations of the AHO model, Eqs.(\ref{Eq21}), (\ref{Eq22}), and (\ref{Eq25}), and those of the 1D $\phi^4$ lattice model, Eqs.(\ref{Eq63}), (\ref{Eq64}), (\ref{Eq69}), and (\ref{Eq70}) can explain the low frequency cut-off in the $\phi^4$ lattice model and gives the cut-off frequency 
\begin{equation}     \label{cutoff}
 \omega_{c}(k) = \sqrt{(2K/m)(1-\cos{k})}.
\end{equation}
The inset of Fig.8(b) compares the observed cut off frequency (symbols) with this equation (line) and perfect agreement is obtained.

\begin{figure}
 \vspace*{-4.5cm}
\begin{center}
  \includegraphics[width=450pt, height=310pt, angle=0]{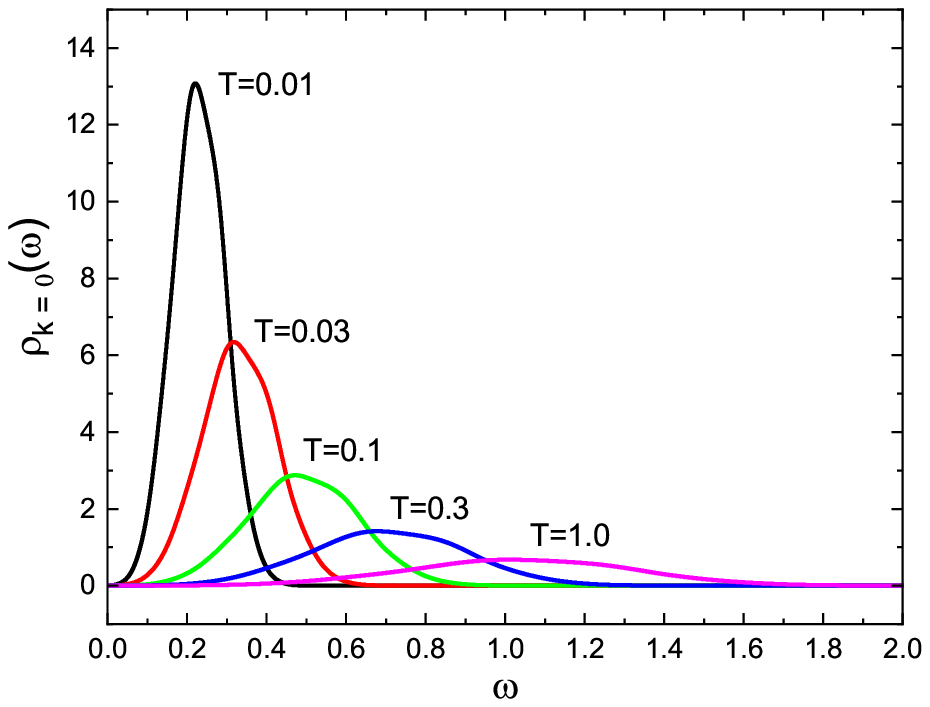}
  \vspace*{-1.0cm}
\end{center}
  \caption{(color online) Spectral function $\rho_{Q_k, Q_k^{\ast}}(\omega)$ at $k=0$ for 1D $\phi^4$ lattice model at different temperatures. Curves are obtained from averaging the results of $400$ random $\{\lambda_i\}$'s with $\Delta=7.0$. Broadening parameter $r=0.3$. Model parameters are $m=1.0$, $K=1.0$, $\gamma=1.0$, and $L=1000$. 
}   \label{Fig9}
\end{figure}

Fig.\ref{Fig9} shows the evolution of the phonon spectral function $\rho_{Q_{k=0}, Q_{k=0}^{\ast}}(\omega)$ with temperature. We used basis dimension $N=20$. For a given $T$, the spectral function curve has a single, almost symmetric peak in the positive frequency regime, located around $\omega = \Omega_{k=0}$, with width $\Gamma_{k=0}$. Both $\Omega_{k=0}$ and $\Gamma_{k=0}$ increases with increasing temperature. We will see from Fig.\ref{Fig12} that both increase as $T^{1/3}$ in the low-temperature regime, while they increase as $T^{1/4}$ in the high-temperature limit. Note that the frequency with the largest $\rho_{Q_{k}, Q_{k}^{\ast}}(\omega)$ is numerically very close to the $\Omega_k$ calculated from Eq.(\ref{Eq48}) from one-shot raw $\rho_{Q_{k}^{\ast}, Q_{k}}(\omega)$ data. Therefore, in this paper, we always use the latter definition for the dispersion.

Comparing $\Omega_k$ to the dispersion $\omega_k$  Eq.(\ref{Eq57}) obtained from a single $Q_k$-basis calculation (equivalent to the $\mathcal{H}_{\nu k}$-expanded basis at $N=1$ as well as to the quadratic variation \cite{Ma2}),
we find the same qualitative behavior and small quantitative deviation. We conclude that the phonon dispersion from $N=1$ PTA is already quite accurate and quantitative improvement of $\Omega_k$ with increasing $N$ is limited.
\begin{figure}
 \vspace*{-2.5cm}
\begin{center}
  \includegraphics[width=450pt, height=340pt, angle=0]{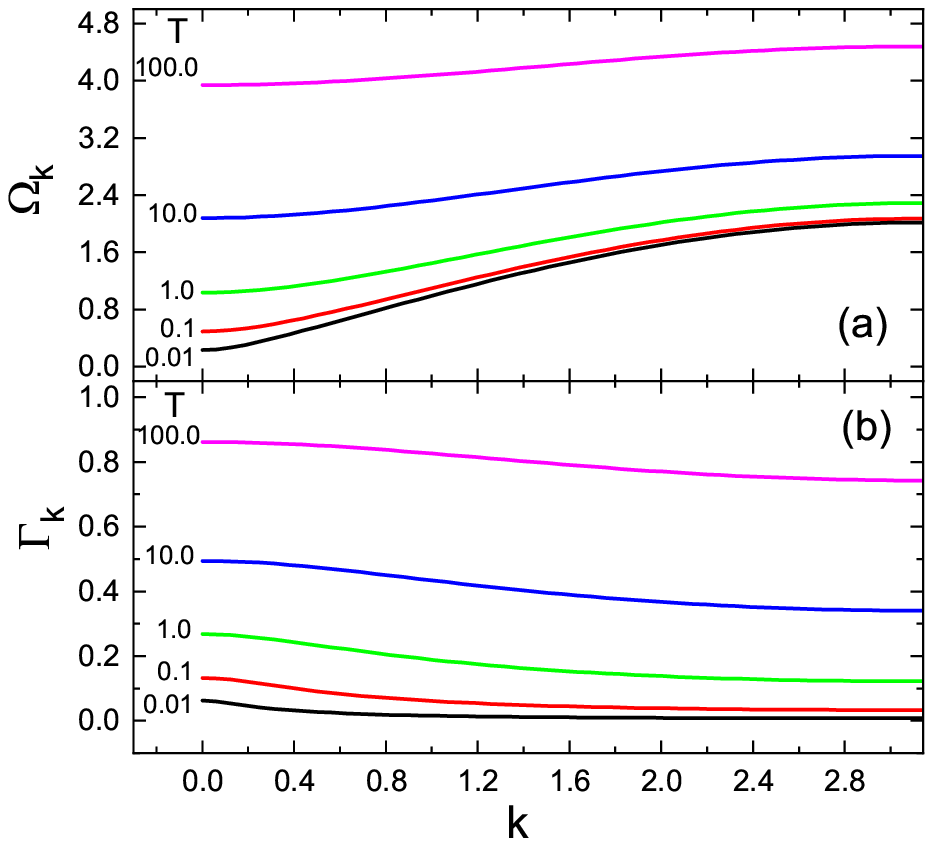}
  \vspace*{-1.0cm}
\end{center}
  \caption{(color online) (a) Dispersion $\Omega_k$ and (b) the spectral width $\Gamma_k$ for 1D $\phi^4$ lattice model. They are extracted from the phonon spectral function $\rho_{Q_{1k}, Q_{1k}}(\omega)$  from PTA on $H_{\nu k}$-expanded basis with $N=20$. Temperatures are marked out in the figure. Model parameters are $m=1.0$, $K=1.0$, $\gamma=1.0$. We use $L=1000$ for $T=0.01$ and $L=100$ for other temperatures.
}   \label{Fig10}
\end{figure}
\begin{figure}
 \vspace*{-3.0cm}
\begin{center}
  \includegraphics[width=410pt, height=320pt, angle=0]{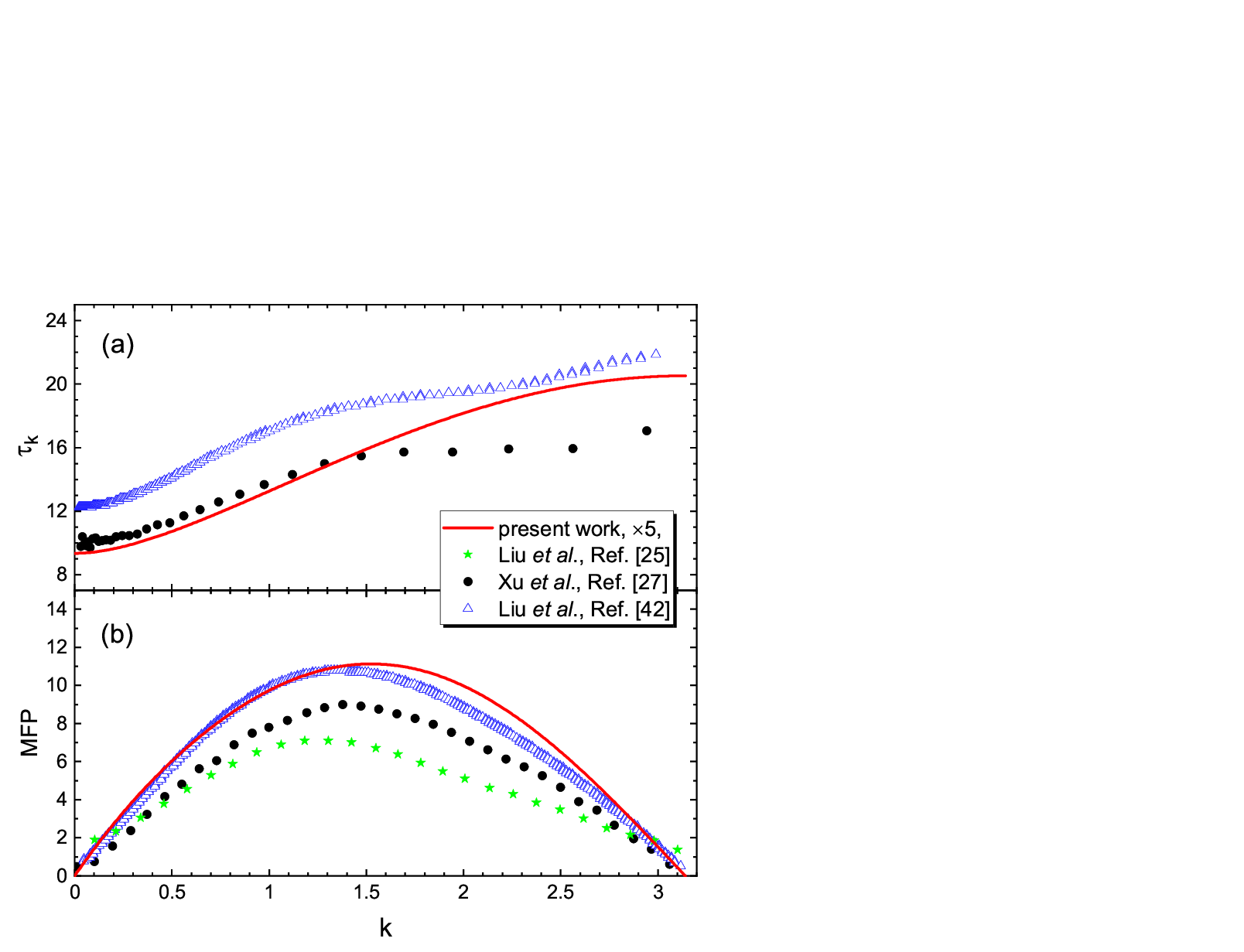}
  \vspace*{-1.0cm}
\end{center}
  \caption{(color online) Comparison of (a) $\tau_k$,  and (b) the mean free path $l_k$ at $T=1.0$ obtained from PTA and various MD simulations in Refs. \cite{Xu2,AP1,YLiu1}. Model parameters are $m=1.0$, $K=1.0$, $\gamma=1.0$, and $L=1000$. PTA results are multiplied by a factor $5.0$ in the figure.
}   \label{Fig11}
\end{figure}

Fig.\ref{Fig10} shows the dispersion $\Omega_k$ and width $\Gamma_k$, extracted from $\rho_{Q_{k}, Q_{k}^{\ast}}(\omega)$ using Eqs.(\ref{Eq48}) and (\ref{Eq49}), respectively. We present them as functions of $k$ for various temperatures. The dispersion $\Omega_k$ shown in Fig.10(a) is quantitatively close to Eq.(\ref{Eq57}), as checked in Fig.\ref{Fig7}(a). The gap at $k=0$ increases with temperature in a two-section power-law fashion, as summarized in Fig.\ref{Fig12}(a). The spectral width shown in Fig.10(b) is a decreasing function of $k$ for all $T$, in agreement with the MD result \cite{YLiu1}. The long wave length phonons that are important to the thermal transport are strongly damped by thermal fluctuations. This leads to a finite thermal conductivity in the thermodynamic limit at $T>0$, in contrast to the lattice models with continuous spatial translational invariance. Note that the data in Fig.\ref{Fig10} converge with respect to increasing chain length $L$. To obtain reliable data for the thermodynamic limit, we used a chain length up to $L=4000$, which is larger than the correlation length of the lowest studied temperature.

In Fig.\ref{Fig11}, we make quantitative comparisons between the results from the PTA and various MD simulations. Fig.\ref{Fig11}(a) shows the phonon life time $\tau_k \equiv 1/(2 \Gamma_k)$ and Fig.\ref{Fig11}(b) shows the mean free path $l_k \equiv \tau_k v_k$ at $T=1.0$. Note that the extraction of $\tau_k$ and $l_{k}$ from MD simulation is a nontrivial task and there are various ways of doing it \cite{AP1,Xu2,Guo1,YLiu1}. This leads to quantitative deviations among different MD results. The PTA results for $\tau_k$ and $l_{k}$ are smaller than those of MD by a factor of $5$ or so. When multiplied by a factor $5$, PTA result has qualitatively the same $k$-dependence with the MD curves. We believe that, apart from the difference in the definition of $\Gamma_k$, the approximation in Eq.(\ref{Eq69}) is a major source for the underestimation of $\tau_k$ and $l_k$ in PTA. In Eq.(\ref{Eq69}), the correlation between $Q_{\mu p}^2$ and $e^{(\lambda_i + \lambda_j)\mathcal{H}_{\nu k}}$ for modes $(\mu p) \neq (\nu k)$ is replaced with those of the same mode. This tends to exaggerate $\langle x^2 e^{(\lambda_i + \lambda_j)\mathcal{H}_{\nu k}} \rangle$ and hence overestimates the broadening $\Gamma_k$ at low temperatures (see Fig.\ref{Fig7}(c)). Further improvement of our theory in this aspect is in progress.

From the benchmark comparison with MD in Fig.7 and Fig.11, we pinpoint certain aspects where the present method of PTA can compete with and has some advantages over MD. For the low temperature limit where the equilibrium state needs to be built in a much longer time, MD is time-consuming. Both the low and high frequency limits of the dynamical correlation function are resource-demanding for MD. The methods of extracting relevant physical quantities are yet to be unified in MD, leading to quantitative deviations among different MD studies, as shown in Fig.11. In contrast, PTA produces results for well-defined quantities within the same order of time for arbitrary temperature and frequency with well-controlled precision.

\begin{figure}
 \vspace*{-3.0cm}
\begin{center}
  \includegraphics[width=440pt, height=300pt, angle=0]{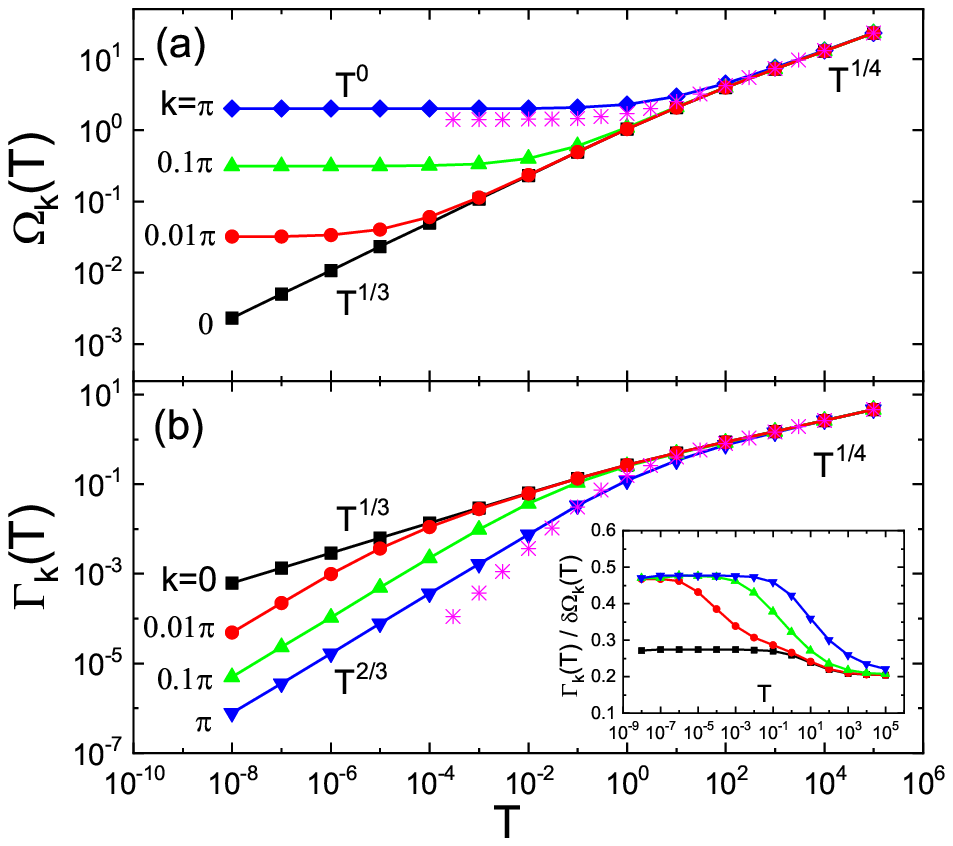}
  \vspace*{-1.0cm}
\end{center}
  \caption{(color online) (a) Dispersions $\Omega_k$ and (b) the spectral widths $\Gamma_k$ as functions of $T$ for various momenta. They are extracted from $\rho_{Q_{k}, Q_{k}^{*}}(\omega)$ of 1D $\phi^{4}$ lattice model, obtained from PTA on $H_{\nu k}$-expanded basis with $N=20$. The low and high temperature asymptotic power laws are marked in the figure.
The stars are results for AHO model. Inset: ratio of $\Gamma_k(T)$ and $\delta \Omega_k(T) = \Omega_k(T) - \sqrt{(2K/m)(1-\cos{k})}$ as functions of $T$ for the same $k$ values as the main figure. Model parameters are $m=1.0$, $K=1.0$, and $\gamma=1.0$. We used $L=100$ (for high $T$) $\sim$ $4000$ (for low $T$) which are much larger than the correlation length at the corresponding temperature. 
}   \label{Fig12}
\end{figure}

Having presented the benchmarking results, below we focus on new physical results obtained by PTA for the 1D $\phi^4$ lattice model in Figs.12, 13, and 14. Fig.\ref{Fig12} shows the temperature dependence of $\Omega_k$ and $\Gamma_k$, for a series $k$ from $0$ to $\pi$. The dispersion $\Omega_k(T)$ shown in Fig.\ref{Fig12}(a) is a three-section power law curve of temperature except at $k=0$. In the high temperature regime $T \gg T_{\text{cr}} \sim 1.0$, $\Omega_k(T) \sim T^{1/4}$ and the curves for all momentum merge into the curve $\Omega(T)$ of the AHO model (stars in the figure). This shows that $T \gg T_{\text{cr}}$ is a local-oscillator regime where oscillations are dominated by local $x^{4}$ potential and are dispersionless. As temperature decreases below $T_{cr}$, $\Omega_k(T) \sim T^{1/3}$, before it finally enter a plateau at very low temperature $T \ll T_{k}$. The crossover temperature $T_{k}$ decreases when $k$ decreases. In particular, $T_{k=0} = 0$ and the intermediate temperature power law $\Omega_{k=0}(T) \sim T^{1/3}$ is thus maintained to $T=0$.

Fig.\ref{Fig12}(b) shows the spectral width $\Gamma_k(T)$. It is also an increasing function with three-section power-law behavior, except at $k=0$. Being different from $\Omega_k$, $\Gamma_k$ decreases with increasing $k$. It tends to zero at $T=0$ for all momentum. At high temperatures $T \gg T_{cr} \sim 1.0$, $\Gamma_k(T) \sim T^{1/4}$ for all momenta and they merge into the $\Gamma(T)$ curve of AHO model (stars). As temperature lowers, $\Gamma_k(T) \sim T^{1/3}$ in a temperature window $ T_{k} \ll T \ll T_{cr}$. The $T_k$ here is the same $k$-dependent crossover temperature in the $\Omega_k(T)$ curve. Finally, when $T \ll T_{k}$, we have the third section with power law $\Gamma_k(T) \sim T^{2/3}$. For $k=0$, $T_{k=0} = 0$. So $\Gamma_{k=0}(T) \sim T^{1/3}$ is maintained to the low temperature limit. As shown in Fig.\ref{Fig7}(c), these power laws are confirmed by the MD results.

The inset of Fig.\ref{Fig12}(b) shows the ratio $\Gamma_k(T)/\delta \Omega_k(T)$ as functions of $T$ for the same $k$'s as in the main figure. Here $\delta\Omega_k(T) \equiv \Omega_k(T) - \Omega_{k}(T=0)$, with $\Omega_{k}(T=0) = \sqrt{(2K/m)(1-\cos{k})}$ being the dispersion at zero temperature. It is seen that the power law $T$-dependence in the numerator and denominator cancels, giving a ratio that is weakly dependent on $T$ and $k$. This is consistent with the asymptotic expressions for $\Omega_k(T)$ and $\Gamma_k(T)$ in Eqs.(\ref{Eq74}) and (\ref{Eq75}).

The peculiar temperature dependence of $\Omega_k(T)$ and $\Gamma_k(T)$ can be understood in terms of the quadratic variation result for $\omega_k$ in Eq.(\ref{Eq57}), by a similar analysis as for the AHO model. It is well known \cite{Boyanovsky1,Ma2,Jia1} that for $\phi^{4}$ model, $\langle x^2 \rangle \sim T^{2/3}$ for $T \ll T_{cr}$ and $\langle x^2 \rangle \sim T^{1/2}$ for $T \gg T_{cr}$. $T_{cr}$ depends on the ratio between $K$ and $\gamma$, and $T_{cr} \sim 1.0$ in our calculation. Putting this into Eq.(\ref{Eq57}), one obtains the asymptotic $T$ dependence of dispersion as
\begin{equation}   \label{Eq74}
 \Omega_{k}(T) \sim 
   \left\{\begin{array}{lllll}
       \Omega_k(T=0) + c \, T^{\frac{2}{3}}, & \,\,\,  (T \ll T_{k} ) \\
        \\
       c^{\prime} \, T^{\frac{1}{3}},      & \,\,\,  (T_{k} \ll T \ll T_{cr})  \\ 
       \\
c^{\prime\prime} \, T^{\frac{1}{4}},    & \,\,\,  (T \gg T_{cr} ).
     \end{array} \right.   
\end{equation}
Here, $T_k$ is obtained by equating $2K(1-\cos{k})$ with $3\gamma \langle x^2 \rangle$ in Eq.(\ref{Eq57}), which gives $T_k \sim \left[(K/\gamma)(1-\cos{k})\right]^{3/2}$. It tends to zero in the limit $k=0$.
Therefore, we call the three regimes of temperature the local oscillator regime ($T \gg T_{cr}$), the pure harmonic regime ($T \ll T_{k}$), where anharmonic interaction does not appear in the dispersion $\Omega_k \approx \omega_k(\gamma=0)$, and the renormalized harmonic regime $T_{k} \ll T \ll T_{cr}$, where the oscillators still form collective waves but with renormalized dispersion since $3\gamma \langle x^2 \rangle \gg \omega_k$ in Eq.(\ref{Eq57}).

For the thermal broadening $\Gamma_k(T)$, we invoke the expression $ \omega_k(E_0) \sim \sqrt{\frac{1}{m} \left[ 2K(1-\cos{k}) + 3 \gamma x_0^{2}\right]}$, according to Eq.(\ref{Eq57}). Here $E_0$ is the initial energy of the mode $k$, which has a fluctuation magnitude of $T$. In the low-temperature limit, $x_0^{2} \sim \langle x^2 \rangle \sim T^{2/3} \sim E_0^{2/3}$. In the high-temperature limit, the energy is dominated by the $x^4$ local potential and we have $x_0^4 \sim E_0$. Inputting these relations into $\omega_k(E_0)$ and taking $E_0 \sim \delta E_0 \sim T$, we obtain the expression
\begin{equation}     \label{Eq75}
 \Gamma_{k}(T) \sim 
   \left\{\begin{array}{lllll}
       d \, T^{\frac{2}{3}}, & \,\,\,  (T \ll T_{k} ) \\
        \\
       d^{\prime} \, T^{\frac{1}{3}},      & \,\,\,  (T_{k} \ll T \ll T_{cr})  \\ 
       \\
d^{\prime\prime} \, T^{\frac{1}{4}},     & \,\,\,  (T \gg T_{cr} ).
     \end{array} \right.   
\end{equation}
Here, we have assumed $T_k \ll T_{cr}$. Eqs.(\ref{Eq74}) and (\ref{Eq75}) summarize the observed results in Fig.\ref{Fig12}.

Eqs.(\ref{Eq74}) and (\ref{Eq75}) suggests a relation between the width $\Gamma_k(T)$ and dispersion $\Omega_k(T)$,
\begin{equation}    \label{relation}
   \Gamma_{k}(T) \approx c \left[ \Omega_k(T) - \Omega_k(0) \right].
\end{equation}
Here, $c$ is weakly dependent on $T$ and $k$. See the inset of Fig.12(b) for its behavior. This relation is in contrast to the assumption of the effective phonon theory \cite{Li1} that $\Gamma_k(T) \approx \epsilon(T) \Omega_k(T)$. Here, $\epsilon(T) = \langle E_{n} \rangle / \langle E_{n} + E_{l} \rangle$, with $E_n$ and $E_{l}$ being the nonlinear and linear potential energies, respectively. The results $\Gamma_k(T)$ and $\Omega_k(T)$ shown in Fig.12 do not agree with this assumption both in the momentum and in the temperature dependences.

\begin{figure}
 \vspace*{-6.5cm}
\begin{center}
  \includegraphics[width=520pt, height=380pt, angle=0]{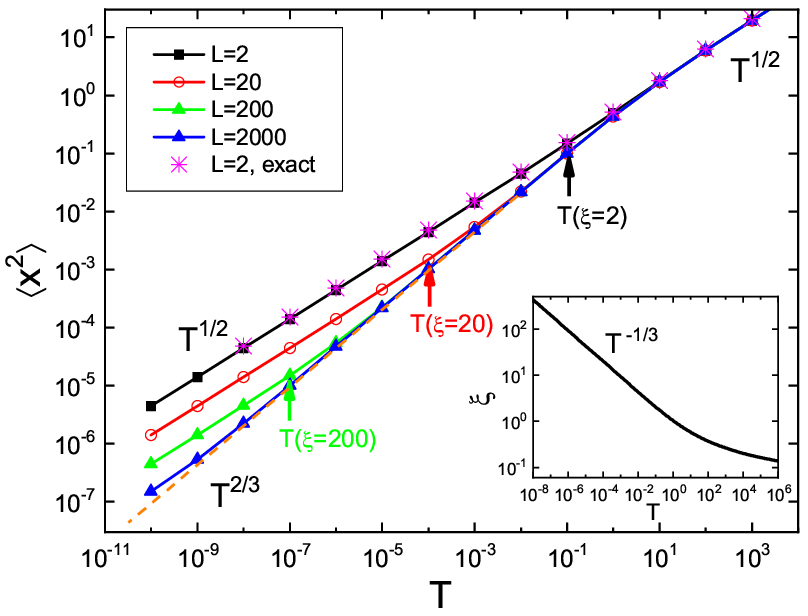}
  \vspace*{-1.0cm}
\end{center}
  \caption{(color online) $\langle x^2 \rangle$ as functions of temperature for various chain length $L$ (symbols with guiding lines). The exact curve for $L=2$ is shown as stars. The low temperature asymptotic powers are marked in the figure. The arrows mark out the temperature at which the correlation length $\xi(T)$ equals to $L$. The dashed line shows $T^{2/3}$ power law. The inset shows the correlation length $\xi$ as a function of $T$ obtained by cluster variation method \cite{Jia1}. We use basis dimension $N=20$ and parameter $\Delta = 7.0$. Model parameters are $m=1.0$, $K=1.0$, and $\gamma=1.0$. 
}   \label{Fig13}
\end{figure}
\begin{figure}
 \vspace*{-3.0cm}
\begin{center}
  \includegraphics[width=380pt, height=270pt, angle=0]{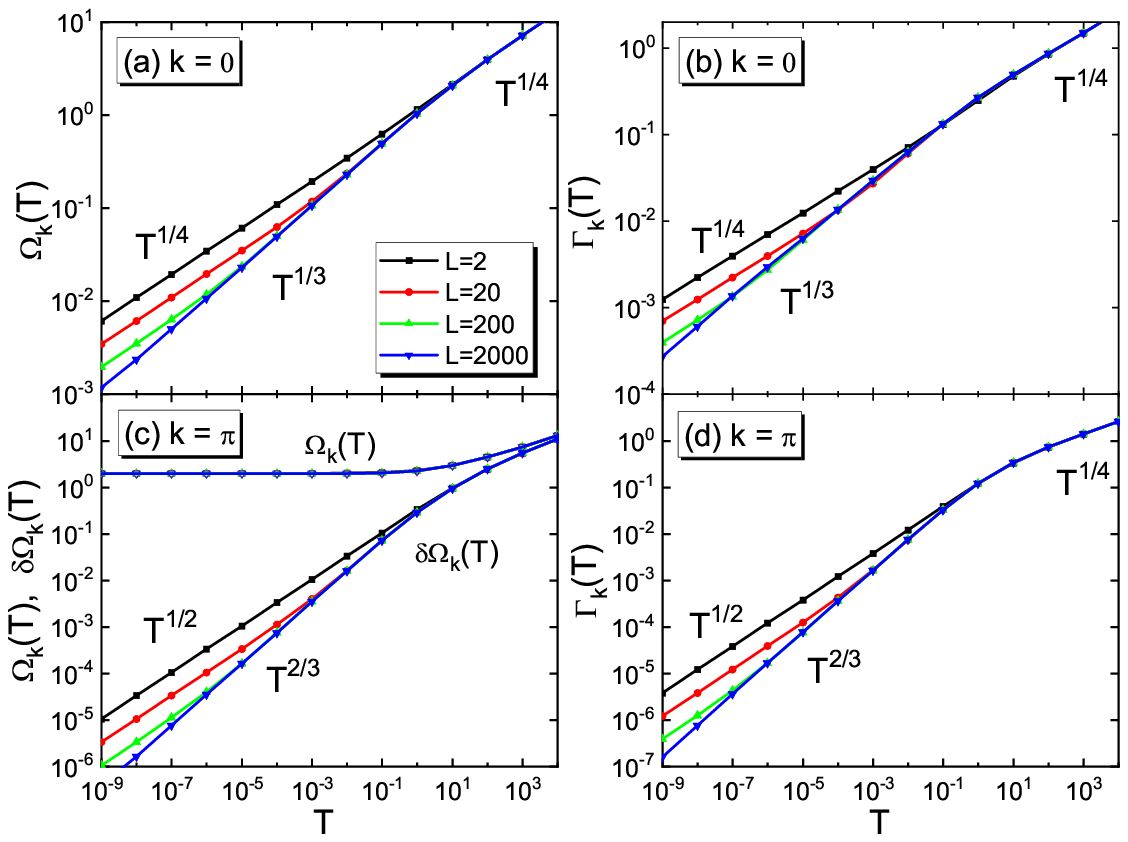}
  \vspace*{-1.0cm}
\end{center}
  \caption{(color online) Dispersion $\Omega_{k}(T)$ and broadening $\Gamma_k(T)$ as functions of temperature for various chain length $L$. (a) $\Omega_{k=0}$, (b) $\Gamma_{k=0}$, (c) $\Omega_{k=\pi}(T)$ (empty symbols) and $\delta \Omega_{k=\pi}(T)=\Omega_{k=\pi}(T) -\Omega_{k=\pi}(0)$ (solid symbols), and (d) $\Gamma_{k=\pi}(T)$. In each panel, from top to bottom, $L=2$, $20$, $200$, and $2000$. Lines are for guiding eyes. We use basis dimension $N=20$ and parameter $\Delta = 7.0$. Model parameters are $m=1.0$, $K=1.0$, and $\gamma=1.0$. 
}   \label{Fig14}
\end{figure}

\end{subsection}

%----------------------------------
\begin{subsection}{short-chain limit of 1D $\phi^4$ lattice model}

So far, we have focused solely on the thermodynamical limit where the chain length $L$ is much larger than the correlation length $\xi(T)$. Our results presented above are obtained with sufficiently large $L$ to avoid any finite-size effect. Here, we report that the $\phi^4$ model has also a short-chain limit characterized by $L \ll \xi(T)$, where physical quantities have distinct power laws from those in the thermodynamical limit. This regime is pertinent to the thermal transport experiments with small size low-dimensional systems, but to our knowledge, has not been discussed in the literature. 

In Fig.{\ref{Fig13}}, $\langle x^2 \rangle(T)$ is plotted for chain size ranging from $L=2$ to $L=2000$. For each $L$, in the high temperature limit $T \gg T_{cr} \approx 1.0$, we have $\langle x^2 \rangle(T) \sim T^{1/2}$, the power law of the atomic limit discussed in previous subsections. As $T$ decreases below $T_{cr}$, $\langle x^2 \rangle \sim T^{2/3}$. Here $T_{cr}$ is the $L$-independent local-nonlocal crossover temperature. Since the correlation length $\xi$ increases with decreasing $T$ as $\xi \sim T^{-1/3}$ in the low temperature limit \cite{Jia1} (see the inset of Fig.{\ref{Fig13}}), when $T$ decreases further and is below a short-chain crossover temperature $T_{sh}$ where $\xi =L$ (marked by the arrows in the figure), $\xi > L$ occurs and $\langle x^2 \rangle(T)$ begins to deviate from the $T^{2/3}$ law. In the low $T$ limit where $L \ll \xi(T)$, $\langle x^2 \rangle \sim T^{1/2}$, being different from the $T^{2/3}$ behavior of the thermodynamical limit. We have checked that the dependence of $T_{sh}$ on $L$ is determined by $\xi(T_{sh})=L$, giving $T_{sh}(L) \sim L^{-3}$, with a coefficient about $1.0$. In the thermodynamical limit, $T_{sh}(L=\infty)=0$ and the $T^{2/3}$ behavior extends to zero temperature. Note that $T_{sh}(L) \leq T_{cr} \approx 1.0$ holds for its largest value $L=2$. This behavior is summarized as
\begin{equation}   \label{avex2_short}
 \langle x^2 \rangle(T) \sim 
   \left\{\begin{array}{lllll}
        T^{\frac{1}{2}}, & \,\,\,  (T \ll T_{sh}(L) ) \\
        \\
       T^{\frac{2}{3}},  & \,\,\,  (T_{sh}(L) \ll T \ll T_{cr})  \\ 
       \\
 T^{\frac{1}{2}},    & \,\,\,  (T \gg T_{cr} ).
     \end{array} \right.   
\end{equation}
This behavior can be understood from the $N=1$ solution, where $\langle x^2 \rangle$ is determined by the self-consistent equations Eqs.(\ref{Eq57}) and (\ref{Eq59}). In the limit of small $L$ and low $T$, the summation of $k$ in Eq.(\ref{Eq59}) is dominated by the $k \sim 0$ terms. The result is therefore similar to that of the atomic limit at $K=0$ or $T \gg T_{cr}$.

A similar short-chain limit exists in the dynamical quantities $\Omega_k(T)$ and $\Gamma_{k}(T)$. In Fig.{\ref{Fig14}}(a) and (b), we show the curves of $\Omega_{k=0}(T)$ and $\Gamma_{k=0}(T)$ for a series $L$. For momentum $k=0$, both quantities have a common three-section power-law behavior. As $T$ decreases, they change from the high temperature atomic-limit behavior $T^{1/4}$, to the non-local long-chain behavior $T^{1/3}$ in  $T_{sh}(L) \ll T \ll T_{cr}$, and to the short-chain behavior $T^{1/4}$ in the low temperature limit $T \ll T_{sh}(L) \leq T_{cr}$. The behavior is summarized as
\begin{equation}   \label{omegak0_short}
\Omega_{k=0}(T), \,\,\, \Gamma_{k=0}(T) \sim 
   \left\{\begin{array}{lllll}
        T^{\frac{1}{4}}, & \,\,\,  (T \ll T_{sh}(L) ) \\
        \\
       T^{\frac{1}{3}},  & \,\,\,  (T_{sh}(L) \ll T \ll T_{cr})  \\ 
       \\
 T^{\frac{1}{4}},    & \,\,\,  (T \gg T_{cr} ).
     \end{array} \right.   
\end{equation}
In Fig.{\ref{Fig14}}(c) and (d), we show the quantities for $k=\pi$. In the low $T$ limit, $\Omega_{k=\pi}(T)$ is a finite constant, not showing the short-chain behaviors. However, the quantities $\delta \Omega_{k=\pi}(T) \equiv \Omega_{k=\pi}(T)-\Omega_{k=\pi}(0)$ and $\Gamma_{k=\pi}(T)$ do have short-chain limit behavior. For other $k >0$, similar behaviors are found and they are summarized as 
\begin{equation}   \label{omegak1_short}
\delta \Omega_{k >0}(T), \,\,\,  \Gamma_{k >0 }(T) \sim 
   \left\{\begin{array}{lllll}
        T^{\frac{1}{2}}, & \,\,\,  (T \ll T_{sh}(L) ) \\
        \\
       T^{\frac{2}{3}},  & \,\,\,  (T_{sh}(L) \ll T \ll T_{cr})  \\ 
       \\
 T^{\frac{1}{4}},    & \,\,\,  (T \gg T_{cr} ).
     \end{array} \right.   
\end{equation}

These results are in principle amenable to experimental tests on low dimensional materials with varying length and temperature. They also show that one must be careful when interpreting the temperature behavior of the experimental or MD data related to thermal transport in low dimensional systems, since at least two energy scales are present to separate the scaling regimes. One is related to the length of chain, and the other is related to the competition between local and non-local interactions. In the present work, we fix parameters $m = K = \gamma=1$ and hence have a definite $T_{cr} \approx 1.0$ and $T_{sh}(L) \approx 1.0 \times L^{-3}$. For a system with arbitrary parameters, the concrete values of the two crossover temperatures will shift according to the exact or approximate scaling properties of the model. Therefore before one can carry out reliable power-law fitting, one must collect data in a sufficiently wide range of temperature to discern the possible crossovers.

\end{subsection}

\end{section}
%========================================================================%%

\begin{section}{Discussions and Summary}

We first discuss the implication of our results for the temperature dependence of thermal conductivity $\kappa(T)$. 
In the Debye formula $\kappa = \sum_{k} C_k v_k l_k$ \cite{Ziman1,Allen1}, $\kappa$ is expressed in terms of the specific heat capacity $C_k = (1/L)\partial \langle H_{k} \rangle / \partial T$, the phonon group velocity $v_k = \partial \Omega_k / \partial k$, and the mean free path $l_k = v_k/ (2\Gamma_k)$.

PTA results for $\Omega_k$ and $\Gamma_k$ can be used to calculate $\kappa$ through the Debye formula. It is known that $C_{k >0}$ ranges between $1.0$ (low $T$) and $0.75$ (high $T$) and is weakly $k$ dependent, and $C_{k=0} = 0.75$ \cite{Ma2}. Approximating $C_k$ as a constant, and inputting Eqs.(\ref{Eq74}) and (\ref{Eq75}) into the Debye formula, we obtain the asymptotic $T$ dependence of $\kappa$ as 
\begin{equation}    \label{Eq84}
 \kappa(T) \sim 
   \left\{\begin{array}{lll}
        T^{- \frac{2}{3}}, & \,\,\,  (T \ll T_{cr}) \\
        \\
        T^{ - \frac{3}{4}},      & \,\,\,  ( T \gg T_{cr}),
     \end{array} \right.   
\end{equation}
with $T_{cr} \approx 1.0$.
Direct numerical evaluation of the Debye formula using PTA input confirms this result. However, this behavior does not agree with the high temperature result $\kappa(T) \sim T^{-1.35}$ from MD simulation in Ref.\cite{Aoki1} and the expression $\kappa(T) \sim T^{-4/3}$ from the effective phonon theory \cite{Li3}. This poses question on the applicability of the Debye formula for describing the thermal conductivity of the $\phi^4$ lattice model. A detailed answer to this question is beyond the scope of this paper and we leave it for a future study.

In this work, we developed the idea of expanding a low level eigen-operator $Q_{\nu k}$ by multiplying it with functions of approximate mode energy $\mathcal{H}_{\nu k}$, to describe the thermal-broadened quasi-particle peak in the spectral function. Here, we have used the basis variable $Q_{\nu k} e^{\lambda_i \mathcal{H}_{\nu k}}$, but it is interesting to know the importance of the inter-mode coupling for the effectiveness of basis. For example, PTA on the more complete basis $Q_{\nu k} e^{\sum_{i \mu p}\lambda^{i}_{\mu p} \mathcal{H}_{\mu p}}$ needs to be investigated to answer this question. We expect that it can give quantitative improvement of $\Gamma_k$. It is also noted that the present $H$-expanded basis does not cover all types of basis operators or scattering processes. For example, the variable $Q_{\nu_1 k_1}Q_{\nu_2 k_2}Q_{\nu_3 k_3}$ ($k_1 + k_2 + k_3 = k \mod{2\pi} )$ with different $k_1 \sim k_3$ is not covered. Scrutinization is therefore needed for applying the present $H$-expansion of the basis to general situations.

For quantum systems, quantum fluctuation is an important source of quasi-particle broadening. In particular, in systems such as frustrated quantum antiferromagnetism or system close to quantum criticality, quantum damping plays a dominant role. The present scheme can be generalized to study the quantum damping of quasi particles or quantum broadening of the quasi-particle peak in the spectral function. A direct application of the present formalism needs additional care, because for quantum systems the conservation identity Eqs.(\ref{A1}) and (\ref{A2}), used to simplify the matrices ${\bf I}$ and ${\bf L}$, must be modified. The GF and the whole formalism of the PTA needs to be replaced with the quantum version. Other than this, there seem to be no other constraints. In case the scattering processes other than those contained in the $H$-expanded basis play a vital role, one must consider the expanding schemes with other (approximately) conserved quantities other than energy, such as the particle number, momentum, spin, {\it etc.}, to describe the decay of quasi-particles due to various fluctuations. 
What we have in mind is the spin-wave damping in frustrated antiferromagnets \cite{Winterfeldt1}. Further study of this interesting subject is in progress.

In summary, we developed the PTA on the $H$-expanded basis to describe the thermal broadening of quasi particles. A zeros-removing technique is used to stabilize the iterative solution of PTA equations. Benchmarking calculations on the classical AHO model and the 1D $\phi^4$ model show that this method can produce semi-quantitative thermal broadening of a quasi-particle peak in the spectral function. Using this method, we disclose the low- and high-temperature power laws of the phonon spectral width $\Gamma_k$ for the 1D $\phi^4$ lattice model, contradicting the assumption in effective phonon theory and raising questions on the applicability of the Debye formula for this model. A short-chain limit of this model is also discovered. Future development and possible extension of this method are discussed.

\end{section}

\vspace*{0.5cm}
\begin{section}{Acknowledgments }

We acknowledge helpful discussions with Y.J. Wu, C.L. Ji, X.G. Ren, L. Wan, Y. Wan, and T. Li. We are grateful to Professor Ren for showing us the method to treat the redundant basis variables. This work is supported by NSFC (Grant No.11974420 and No.12075316). 

\end{section}

\appendix{}

\section{Derivation of ${\bf I}$ and ${\bf L }$ matrices for AHO model on bases $B_2$ and $B_3$ }

In this Appendix, we present the derivation of ${\bf I}$ and ${\bf L }$ matrices for AHO on the bases $B_2$ and $B_3$. 

Before presenting the details, we present the identity to be used for simplifying the expressions \cite{Ma2}. For arbitrary variables $A$ and $B$, we have
\begin{equation}  \label{A1}
   \langle \{A, B\} \rangle = \beta \langle \{A, H\} B \rangle,
\end{equation}
and for arbitrary number $\theta$,
\begin{equation}   \label{A2}
  \langle \{A, H\} B e^{\theta H} \rangle = \frac{1}{\beta -\theta} \langle \{A, B\} e^{\theta H} \rangle.
\end{equation}
Eq.(\ref{A1}) is proved in Appendix B. Eq.(\ref{A2}) is obtained by letting $K=1$, $X_1= H$, and $g(X_1) = e^{\theta H}$ in the identity Eq.(\ref{B3}).

%-----------------------------------------%
\begin{subsection}{Basis $B_2$}

For basis $B_2 = \{ A_i = xf_i \}$ ($f_i = e^{\lambda_i H}$), the matrix elements $I_{ij} = (xf_i | xf_j )$ is calculated through Eq.(\ref{A1}) as
\begin{eqnarray}   \label{A3}
  I_{ij} &=& \langle \{ xf_i, \{xf_j, H\} \} \rangle  \nonumber \\
     &=& \beta \langle \{x, H\} \{x, H\} f_i f_j \rangle.
\end{eqnarray}
Using Eq.(\ref{A2}) (letting $A=x$, $B=\{x, H\}$, and $\theta = \lambda_i+\lambda_j$), and $\{x, \{x, H\}\} = 1/\mu$, we obtain
\begin{eqnarray}   \label{A4}
  I_{ij} &=& \langle \{x, \{x, H\} \} f_i f_j \rangle + (\lambda_i + \lambda_j)\langle \{x, H\} \{x, H\} f_i f_j \rangle   \nonumber \\
  &=& \frac{\beta}{\mu \left(\beta - \lambda_i - \lambda_j \right)} \langle f_i f_j \rangle.
\end{eqnarray}

The matrix element $L_{ij}= -(xf_i |\{ \{xf_j, H\}, H\} )$ is calculated similarly as
\begin{eqnarray}   \label{A5}
L_{ij} &=& -\langle \{ xf_i, \{ \{ \{x, H\}, H\}, H\} f_j \} \rangle  \nonumber \\
     &=& -\beta \langle \{x, H\} \{ \{ \{x, H\}, H\}, H\} f_i f_j \rangle.
\end{eqnarray}
Using Eq.(\ref{A2}) (letting $A=x$, $B=\{ \{ \{x, H\}, H\}, H\}$, and $\theta = \lambda_i+\lambda_j$) and $\{x, \{ \{ \{x,H\}, H\}, H\} \} = -(1/\mu^{2}) \left(\mu \omega_0^2 + 12 \alpha x^2 \right)$, we obtain
\begin{eqnarray}   \label{A6}
  L_{ij} &=& \frac{\beta}{\mu^2 \left(\beta - \lambda_i - \lambda_j \right)} \left[ \mu\omega_0^2 \langle f_i f_j \rangle + 12 \alpha \langle x^2 f_if_j \rangle \right].   \nonumber \\
  &&
\end{eqnarray}
Eqs.(\ref{A4}) and (\ref{A6}) are Eq.(\ref{Eq21}) in the main text.

\end{subsection}
%-------------------------------------

\begin{subsection}{Basis $B_3$}

For basis $B_3$, the basis variables are denoted in Eq.(\ref{Eq26}) as $A_{\mu i} = a_{\mu} f_i$, with $a_{1} = x$, $a_{2} = x^3$, and $f_i = e^{\lambda_i H}$. The matrix elements of ${\bf I}$ and ${\bf L}$ read
\begin{equation}   \label{A7}
I_{\mu \nu}^{ij} = \langle \{ A_{\mu i}, \{A_{\nu j}, H\} \} \rangle,
\end{equation}
and
\begin{equation}   \label{A8}
L_{\mu \nu}^{ij} = - \langle \{ A_{\mu i}, \{ \{ \{A_{\nu j}, H\}, H\}, H\} \}\rangle,
\end{equation}
respectively. 
Using the same method used for basis $B_2$ above, we obtain
\begin{eqnarray}   \label{A9}
I_{\mu \nu}^{ij} &=& \frac{\beta}{\beta - \lambda_{i} - \lambda_{j}} \langle I_{\mu \nu}^{(0)} f_{i}(H) f_{j}(H) \rangle,  \nonumber\\
L_{\mu \nu}^{ij} &=& \frac{\beta}{\beta - \lambda_{i} - \lambda_{j}} \langle L_{\mu \nu}^{(0)} f_{i}(H) f_{j}(H) \rangle.
\end{eqnarray}
Here, the variables $I^{(0)}_{\mu \nu} = \{a_{\mu}, \{a_{\nu}, H\} \}$ and $L^{(0)}_{\mu \nu} = -\{a_{\mu}, \{ \{ \{a_{\nu}, H\}, H\}, H\} \}$.
Calculation gives
\begin{eqnarray}   \label{A10}
 &&  I_{11}^{(0)} = \frac{1}{\mu},   \nonumber \\
 &&  I_{12}^{(0)} = I^{(0)}_{21} = \frac{3}{\mu} x^2,   \nonumber \\
 &&  I_{22}^{(0)} = \frac{9}{\mu} x^4,
\end{eqnarray}
and
\begin{eqnarray}   \label{A11}
  && L_{11}^{(0)} = \frac{1}{\mu^2} V^{{\prime}{\prime}}(x),   \nonumber \\
  && L_{12}^{(0)} = -\frac{18}{\mu^3}p^2 + \frac{18}{\mu^2}x V^{\prime}(x) + \frac{3}{\mu^2} x^2 V^{\prime \prime}(x),   \nonumber \\  
  && L_{21}^{(0)}  = \frac{3}{\mu^2} x^2 V^{\prime \prime}(x),  \nonumber \\
   && L_{22}^{(0)} = -\frac{54}{\mu^3} x^2 p^2 + \frac{54}{\mu^2}x^3 V^{\prime}(x) + \frac{9}{\mu^2} x^4 V^{\prime \prime}(x).   \nonumber \\
   &&
\end{eqnarray}
Here $V(x)$ is the potential function Eq.(\ref{Vx}). Inserting these equations into Eq.(\ref{A9}), we get the expressions for $I_{\mu\nu}^{ij}$ and $L_{\mu \nu}^{ij}$. They contain the averages $\langle f_i f_j \rangle$, $\langle x^2 f_i f_j \rangle$, $\langle x^4 f_i f_j \rangle$, $\langle p^2 f_i f_j \rangle$, and $\langle x^2 p^2 f_i f_j \rangle$. The first three can be calculated self-consistently. The last two can be simplified by the conservation identity Eq.(\ref{A2}).

In Eq.(\ref{A2}), we let $A= x$, $B=p$, and $\theta = \lambda_i + \lambda_j$, we obtain
\begin{equation}    \label{A12}
    \langle p^2 f_i f_j \rangle = \frac{\mu}{\beta - \lambda_i - \lambda_j} \langle f_i f_j \rangle.
\end{equation}
Similarly, letting $A=x^2$ and $B=xp$, we obtain
\begin{equation}    \label{A13}
    \langle x^2 p^2 f_i f_j \rangle = \frac{\mu}{\beta - \lambda_i - \lambda_j} \langle x^2 f_i f_j \rangle.
\end{equation}
Using Eqs.(\ref{A12}) and (\ref{A13}), we finally obtain Eqs.(\ref{Eq27}) and (\ref{Eq28}). Note that the conservation relation Eq.(\ref{A2}) is also used to reduce $L^{ij}_{12}$ to $L^{ij}_{21}$. The final result for ${\bf L}$ is Hermitian.

\end{subsection}

\section{Proof of a conservation identity }

In this Appendix, we derive an identity to be used in the derivation of expressions for ${\bf I}$ and ${\bf L}$ in Appendix A. 

First, we cite the conservation identity Eq.(15) of Ref.\cite{Ma2}, {\it i.e.}, for arbitrary variables $A$ and $B$, we have
\begin{eqnarray}  \label{B1}
   \langle \{A(t), B(t^{\prime} )\} \rangle &=& \beta \langle \{A, H\}(t) B(t^{\prime}) \rangle \nonumber \\
   &=&  - \beta \langle A(t)\{ B, H\}(t^{\prime}) \rangle. 
\end{eqnarray}
In the above equation, the average is defined as the canonical ensemble average in an equilibrium state at temperature $T$. $\beta = 1/k_B T$ is the inverse temperature.
In the equal time situation $t=t^{\prime}$, we get
\begin{eqnarray}  \label{B2}
   \langle \{A, B\} \rangle &=& \beta \langle \{A, H\} B\rangle \nonumber \\
   &=&  - \beta \langle A \{ B, H\} \rangle. 
\end{eqnarray}
In the first equation of Eq.(\ref{B2}), we replace $B$ with $B g(X_1, X_2, ..., X_K)$. Here $X_i$ is a conserved quantity satisfying $\{ X_i, H\} =0$ ($i=1,2,..., K$) and $g$ is an arbitrary $K$-variable function. We get the following conservation identity,
\begin{eqnarray}   \label{B3}
&&   \langle \{A, B\} g(X_1, X_2, ..., X_K) \rangle \nonumber \\
& = & \beta \langle \{A, H\} B g(X_1, X_2, ..., X_K)  \rangle  \nonumber \\
&& - \sum_{i=1}^{K} \langle \{A, X_i\} B \frac{\partial }{\partial X_i} g(X_1, X_2, ..., X_K) \rangle.
\end{eqnarray}
Eq.(\ref{A2}) is a special case of Eq.(\ref{B3}), with $K=1$, $X_1 =H$, and $g(X_1) = e^{\theta H}$.
%--------------------------------------------------------------

\end{document}